# Multislice Hollow Ptychography for Simultaneous Atomic-Layer-Resolved 3D Structural Imaging and Spectroscopy

Yu Lei[1] and Peng Wang[1,2*]

[1] Department of Physics, University of Warwick, Coventry CV4 7AL, UK

[2] Electron Microscopy Research Technology Platform (EM-RTP), University of Warwick, Coventry CV4 7AL, UK

* Correspondence to: peng.wang.3@warwick.ac.uk

## Abstract:

Electron–matter interactions in electron microscopy produce both elastic and inelastic scattering, forming the basis for imaging and spectroscopy. However, the integration of electron energy loss spectroscopy (EELS) with 4D-STEM and electron ptychography remains challenging because of detector geometry conflicts. Song et al. solved this issue by introducing a hollow-type pixelated detector that enables hollow ptychography and allows low-angle electrons to go through to the EELS spectrometer. The single-slice approach of hollow ptychography proves effective for 2D thin materials but struggles with multiple scattering in thicker specimens. Here, we introduce multislice hollow ptychography (MHP), a robust imaging modality that overcomes these limitations by accounting for multiple scattering. MHP enables high-resolution structural imaging from hollow diffraction patterns while remaining compatible with simultaneous EELS acquisition. It potentially can provide sub-angstrom lateral resolution at intermediate doses ($10^4$–$10^8$ $e^-/Å^2$) and supports full 3D atomic-layer reconstruction at ultrahigh doses ($\geq 10^8$ $e^-/Å^2$) at the resolution limit set by lattice vibrations and with up to 70% of total electrons available for spectroscopy. This flexible framework facilitates correlative 3D imaging and chemical mapping in complex materials, including energy and quantum materials.

## 1. Introduction

The fundamental principle of electron microscopy depends on electron-material interactions which result in elastic scattering that changes electron paths without energy loss or inelastic scattering that transfers energy to material excitations including phonons, plasmons and electronic transitions. Scanning transmission electron microscopy (STEM), when integrated with electron energy loss spectroscopy (EELS), utilizes both scattering channels to enable atomic-scale characterization of complex materials[1-4]. STEM High-angle annular dark field (STEM-HAADF) imaging detects elastically scattered electrons at high angles, producing Z-contrast images that are particularly sensitive to heavy elements[1,5]. Simultaneously, inelastically scattered electrons are energy-filtered by a spectrometer to provide rich spectroscopic information on local composition, bonding, and oxidation states[6]. This dual-modality capability makes STEM-HAADF and EELS one of the most powerful and widely implemented techniques for probing the structural and electronic properties of technologically important systems such as functional oxides[3,4] and battery materials[7,8] .

STEM-HAADF imaging provides great advantages because of its incoherent imaging properties, including intuitive Z-contrast and robustness to dynamic scattering, but it has several intrinsic limitations. Integration over a high-angle annular detector eliminates phase information, which reduces its sensitivity to image light elements and weak scatterers[1,5]. This technique has low beam dose efficiency, because only a small percentage of incident electrons scattered at high angle effectively contribute to image contrast, which restricts its application for dose-sensitive materials. The depth resolution of HAADF optical sectioning depends on the depth of focus, which extends between 10–30 Å based on probe convergence angle, thus making true atomic-layer-resolved depth resolution challenging[9,10]. 3D atomic resolution imaging relies on tilt-series tomography[11], which also substantially increases the total electron dose.

The recent development of detector technology[12,13] has led to 4D-STEM[14], which records full diffraction pattern throughout a two-dimensional raster of probe positions. Through this mode, one can virtually traditional imaging modes including bright field (BF), annular bright field (ABF), HAADF and differential phase contrast imaging (DPC)[15], as well as access powerful computational imaging techniques such as electron ptychography[16,17]. Unlike HAADF, ptychography reconstructs both amplitude and phase of the specimen complex transmission function using the interference encoded in overlapping diffraction patterns. This method experimentally provides superior spatial resolution beyond the conventional diffraction limit diffraction limit of the lenses[18,19], enhanced sensitivity to both light and heavy elements[20-23], and improved dose efficiency[24-27], which benefits wider applications of characterizations of materials[28,29] and biological samples[25,30]. Furthermore, multislice ptychography[31] offers an advanced approach by explicitly modeling the depth-dependent evolution of the electron probe and its interaction with the sample. A proof-of-principle demonstration of multislice electron ptychography[32] show that by accounting for electron wave propagation through thick specimens, it can suppress the dynamical effects that limit single-slice approaches (ePIE[33,34], WDD[17,22]). Latest work by Chen et al[35]. demonstrated that it can achieve lateral atomic-resolution limits set by lattice vibrations and an improved depth resolution of 3.9 nm in thick, strong scattering specimens, facilitating 3D reconstruction of structural variations[36-40].

These capabilities of 4D-STEM ptychography make it an attractive alternative for traditional STEM-HAADF and EELS workflows, particularly in studies requiring dose efficiency or light-atoms imaging. However, the main obstacle in this application is its incompatibility with EELS acquisition because conventional ptychographic phase retrieval requires capturing the full diffraction pattern, including the intense low angle scattered electrons. However, the same low-angle electrons are typically needed for EELS, as they carry the inelastically scattered signal into the spectrometer. In conventional configurations, the fast pixelated detector[12,13], which records the entire electron diffractions, blocks the entrance to the EELS spectrometer, thus preventing simultaneous structural and spectroscopic data acquisition. This fundamental competition between imaging and spectroscopy for the same portion of the beam limits the integration of ptychography into STEM-EELS workflows.

To address this limitation, a new 4D-STEM microscope configuration with a hollow-type pixelated detector, as shown in Fig. 1a. was proposed by Song et al.[41], which enables the so called hollow ptychography to emerge as a promising strategy. This hollow detector excludes the low-angle region of the bright field disk in the diffraction pattern and redirects low-angle scattered electrons to the EELS spectrometer, while retaining high-angle region for various other types of structural imaging modalities, such as HAADF and DPC. When combined with ptychography, they demonstrated that high-resolution phase reconstructions remain robust for 2D monolayer $MoS_2$, even with the exclusion of low-angle electrons using single-slice reconstruction algorithm. By varying the semi-angle of the hollow region, they showed that subatomic resolution (e.g., resolving the 91 pm). This work established a practical dose-partitioning strategy between 4D-STEM imaging and spectroscopy. However, contrast reversal effects in thicker $MoS_2$ samples were observed, attributed to multiple scattering[41]. This suggests the limitations of single-slice hollow ptychography (SHP) in strongly scattering systems. inherently limited in their ability to account for wave propagation through the sample thickness, resulting in degraded resolution and contrast and lack of 3D structural information.

Here, we develop new imaging modality and associated algorithm, termed multislice hollow ptychography (MHP), based upon multislice approach[42], which enables accurate phase reconstruction from hollow-type diffraction patterns with a large fraction of the central beam excluded. We further investigate if hollow ptychography can handle multiple scattering with this approach (Fig. 1b). Using experimental ptychographic dataset of a strongly scattering perovskite oxide, 21 nm thick $PrScO_3$ (PSO) shared in the literature[35] together with our simulations, we systematically evaluate both achievable lateral and depth resolution under different hollow semi-angles (HSAs), total electron dose and thermal lattice vibration conditions. The simulation results show that multislice hollow ptychography (MHP) enables adjustable control over lateral resolution, depth sectioning, and compatibility with EELS, which makes MHP a powerful platform for correlative 3D atomic-layer-resolved imaging and spectroscopy of complex materials.

## 2. Results

**Robustness High-resolution Imaging for Strongly Scattering Materials**

PSO, a complex perovskite oxide, previously studied in the work[35] presents a challenging system for assessing the effectiveness of advanced ptychography algorithms under conditions of strong multiple scattering. In this study, we reused the same 4D-STEM data[35] from a 20nm [001]-orientated PSO crystal to exam to hollow ptychography by progressively masking the central diffraction region, systematically increasing the HSA, from 0, 0.5α, 0.75α and 0.95α (Fig. 1d-g) to assess the balance between structural imaging capability and the potential for simultaneous EELS mapping. The reconstruction method in this work is based on an iterative Least Squares Maximum Likelihood (LSQML) method, modified from the PtychoShelves framework[43]. Details of the dataset and reconstruction method can be found in Supplementary Note 1 and 2, respectively.

SHP reconstructions revealed its limitations in these thicker specimens consistent with previous work[41]. As the HSA increased, the reconstructed phase images (Fig. 1h-k) and their corresponding power spectra (Supplementary Fig. 2), respectively displayed decreasing contrast, feature blurring, and inaccurate localization of atomic columns. Notably, closely spaced Pr atomic pairs (Fig. 1c and Supplementary Fig. 1) became indistinguishable and merged into single, large intensity peaks, while Sc atoms could not be clearly resolved from surrounding oxygen atoms of configuration 2 labeled as O(2) in Fig. 1c at HSAs beyond 0.5α. The oxygen columns of configuration 1, labeled as O(1) in Fig. 1c located between Pr and Sc sites became increasingly obscured due to the enlarged apparent radii and inaccurate intensities of the adjacent heavier atoms. The quantitative trends in lateral resolution, plotted in Fig. 1p, show that beyond 0.5α hollow radius, the lateral resolution in SHP reconstructions rapidly deteriorates, ultimately exceeding 1Å.

In contrast, the MHP reconstructions maintained high spatial frequency information even as the HSA increased. The MHP reconstructions (Fig. 1l-o) and their corresponding power spectra (Supplementary Fig. 3) showed that even at large HSAs (up to 0.75α), the atomic lattice of PSO remained well-resolved and accurately reconstructed. The atomic dumbbells of Pr, previously merged in the SHP reconstructions, were clearly resolved as distinct atoms. Sc atoms were separated from surrounding oxygen, O(2), and oxygen columns, O(1) located between Pr and Sc were better resolved, exhibiting enhanced contrast and more accurate localization. These images showed excellent agreement with the expected x-y atomic structure model (Fig. 1c), resolving fine details that single-slice reconstructions failed to preserve. The resolution trend in Fig. 1(p) confirms that in the multislice approach, lateral resolution degrades only slightly up to ~0.75α, with more significant decline beyond ~0.90α. Notably, MHP consistently outperforms SHP across all HSAs, despite reduced data redundancy (Supplementary Fig. 4b) caused by the missing central region in the diffraction patterns (Supplementary Fig. 3a-d). These findings demonstrate for the first time that, despite the partial absence of central beam information, hollow ptychography can still be effectively integrated with the multislice approach, offering a powerful framework to overcome resolution degradation caused by multiple scattering in thick specimens. The ability of the multislice algorithm to accommodate large

HSAs up to 0.75–0.90α, while maintaining atomic resolution highlights its suitability for correlative structural and chemical mapping in complex materials systems like transition metal oxides[2-4,44], where simultaneous EELS mapping and high-resolution structural imaging are largely required.

**Dose-dependent MHP Reconstructions**

Understanding electron dose dependence is important for balancing lateral resolution, noise, and potential beam-induced damage, when performing simultaneous EELS mapping, because the total dose must be divided between ptychographic imaging and spectroscopic acquisition. To investigate how electron dose influences on MHP reconstructions, simulations of 4D-STEM datasets from a 20 nm PSO crystal slab were performed using abTEM[45] at 300 kV with a convergence semi-angle of 21.4 mrad, under total electron dose conditions of $10^5$, $10^4$, and $10^3$ e/Å². HSAs were set at 0, 0.5α, 0.75α, and 0.95α. Details of the simulation parameters are provided in Supplementary Note 3.

The reconstructed phases (Fig. 2) and corresponding power spectra (Supplementary Fig. 5) show how the performance of MHP reconstructions across these dose regimes. At a dose of $10^5$ e/Å² or above (Fig. 2a-d), reconstructions exhibit strong, sharp phase contrast of all atomic columns, including challenging oxygen sites between Pr and Sc. High-order reflections, such as are consistently recovered in power spectra (Supplementary Fig. 5a-d), and even at large HSAs up to 0.95α, the atomic lattice remains fully resolved. At $10^4$ e/Å² (Fig. 2e-h), reconstructions maintain good phase contrast and resolution (Supplementary Fig. 5e-g) up to 0.75α HSA. Fine details such as closely spaced Pr atomic dumbbells and oxygen columns are clearly recovered, although slight degradation in contrast is observed beyond this HSA. These intermediate dose reconstructions highlight a practical balance point, delivering excellent structural resolution while minimizing total electron exposure. At the lowest dose of $10^3$ e/Å² (Fig. 2i-l), reconstructions show reduced contrast and increased noise, particularly at HSAs beyond 0.5α, where heavier atomic columns become blurred and oxygen columns are difficult to resolve.

These dose-dependent trends in lateral resolution are quantitatively summarized in Fig. 2m, which plotted resolution as a function of HSAs (bottom x-axis) and the corresponding percentage of total electrons contributing to EELS (top x-axis). During the estimation of EELS fraction, the electrons scattered outside of the bright-field dick were considered. At the doses of $10^5$ or above, lateral resolution remains consistently at sub-angstrom levels, even up to an HSA of 0.95α, which corresponds to 70% of the total electrons contributing to EELS. At $10^4$ e/Å², the resolution remains around 0.7 Å up to approximately 0.9α HSA (corresponds to 75% to EELS), with only moderate degradation at larger HSA. However, at the low-dose condition of $10^3$ e/Å², lateral resolution of 1 Å deteriorates rapidly beyond 0.6α (approximately 35% of the total electrons to EELS), ultimately exceeding 2 Å at the largest hollow regions. These results highlight the flexibility of MHP, enabling tailored trade-offs between imaging resolution and EELS signal strength. The approach is thus well-suited for general-purpose, dose-efficient 2D structural and chemical mapping across a broad spectrum of material systems[46,47].

**Lattice-vibration-dependent 3D MHP Optical Sectioning of a Crystal**

The capability to resolve the 3D spatial arrangement of atoms within materials along the beam direction has been a long-standing goal for optical sectioning techniques in STEM, such as HAADF[9,10] and more recently, ptychography[32,35]. To evaluate the intrinsic depth resolution of MHP, we first conducted simulations using a single Pr atom in vacuum without thermal lattice vibrations (TLV) under various convergence semi-angles (CSAs) and HSAs, while keeping the collection angle and total dose constant (Details are provided in Supplementary Note 5 and Supplementary Fig. 6). The results show that increasing CSA from 20 to 80 mrad significantly improves depth resolution, from ~9.5 Å to ~2.8 Å, and that the resolutions remain stable across HSAs up to ~0.95α. These findings confirm that depth resolution in MHP is primarily determined by CSA and is largely independent of the hollow geometry under typical conditions.

The feasibility of atomic-layer 3D reconstruction of a crystalline sample in different TLV conditions was investigated using MHP at a large CSA. To evaluate the MHP performance under strong scattering conditions, simulations were conducted on a [001]-oriented PSO crystal. In this orientation, the distorted perovskite oxide has alternating layers of Pr, Sc, and O atoms along the beam direction (Supplementary Fig. 1). The interplanar distances vary significantly: ~8 Å for Pr–Pr, ~4 Å for Sc–Sc, and both ~9 Å and ~3 Å for two distinct configurations of O–O layers, labeled O(1) and O(2), respectively (Supplementary Fig. 1d-f). These spacings and the type of atoms set critical thresholds for depth resolution. To meet this criterion, a large probe CSA of 80 mrad capable of achieving a theoretical resolution of 2.8 Å (Supplementary Fig. 6o) was used. Thermal atomic motion at 300 K (room temperature), 90 K (liquid nitrogen) and 4 K (liquid helium) was accounted for using the frozen phonon approach[48,49]. Details of simulation parameters are provided in Supplementary Note 6.

To investigate the influence of TLV on MHP reconstructions, we compared the reconstructed phase at three Z slices from a PSO crystal at temperatures of 300 K (room temperature), 90 K (liquid nitrogen), and 4 K (liquid helium) using the full diffraction pattern (HAS = 0 α) at a dose of $10^8$ $e^-/Å^2$ (Fig. 3 a-i). At 300K, only atomic columns with a large distance of 8~9 Å, such as Pr (Fig. 3a) and O(1) (Fig.3b) can be clearly resolved. The heavier Sc columns with a 4 Å spacing become indistinguishable, as indicated by the profile (—) in Fig. 3k. For the O(2) columns, the O-O dumbbell pairs separated by a 5 Å are visible and individual O atoms with the dumbbell are not resolved (Fig. 3c and m).

At 90K (Fig. 3d-f) and 4K (Fig. 3g-i), all atomic layers, Pr, Sc and both configurations of O(1) and O(2), become clearly resolved along the depth direction. The line profiles in Fig. 3j-m show that as temperature decreases, reduced TLV strongly enhance ptychographic phase signal and depth sectioning performance. This improvement attributes to the fact that atomic vibrations are suppressed at lower temperature, leading to a sharper and tighter atomic potential. This greatly enhances the coherence of electron scattering and results in stronger, more localized phase signals. Therefore, reduced TLV at cryogenic temperatures directly improves the quality of 3D optical sectioning and enable atomic-layer 3D reconstruction of all atomic species including light atoms such as oxygen, using multislice ptychography.

**Dose-dependent 3D MHP Optical Sectioning of a $PrScO_3$ Crystal under Cryogenic Conditions**

Reconstructions were carried out by varying the HSA from 0 to 0.95α and examining the effects of four different electron doses: $10^8$, $10^7$, $10^6$, and $10^5$ $e^-/Å^2$ (Supplementary Fig. 9-10) at 90 K as an example, although slightly better performance can be expected under 4 K conditions for PSO. At the highest dose of $10^8$ $e^-/Å^2$ or above reconstructions remained robust across the full HSA range (Supplementary Fig. 9 a-l). All atomic layers, Pr, Sc and both configurations of O(1) and O(2), were clearly resolved in depth direction, even when the central beam was excluded up to HSA = 0.97α (Fig. 4d-f). The Pr and Sc columns showed strong peaks, while both oxygen configurations showed distinguishable depth features despite their lower scattering strengths (Supplementary Fig. 13a-d). At $10^7$ $e^-/Å^2$ (Supplementary Fig. 9 m-x), reconstructions of the Pr columns, as well as O (1) with its wide 9 Å spacing, remained stable for HSAs up to 0.75α (Fig. 4 j-k). However, the Sc and O (2) columns (Fig. 4k-l), with a much narrower spacing of ~3-4 Å, became increasingly difficult to resolve. Even with the full diffraction pattern (HSA = 0 as shown Fig. 4i), the weak scattering signal and tight depth packing prevented these layers from being recovered as distinct maxima (Supplementary Fig. 13e-h). The loss of resolution in this case suggests the limits imposed by both signal-to-noise ratio and the intrinsic depth response width of the reconstruction. The Pr and O (1) layers remained resolvable up to HSA = 0.75α (Fig. 4p-q) at $10^6$ $e^-/Å^2$, due to their large 8 Å separation. By the dose down to $10^5$ $e^-/Å^2$, only the Pr atomic layers remained resolvable for HSAs up to 0.5α (Fig. 4v). Their strong scattering signal and wide interplanar spacing allowed them to be sufficiently distinguishable along the beam direction, even under dose-limited and hollow conditions. Sc and both O configurations were no longer distinguishable (Fig. 4w-x). Phase retrieval under these conditions produced noisy and elongated features, with no clear maxima corresponding to atomic columns in the depth profiles (Supplementary Fig. 13n-p). The results show that atomic-layer depth sectioning with MHP depends on multiple factors: probe CSA, HSA, total electron dose, atomic number, and interplanar spacing. At a sufficiently large CSA (e.g. 80 mrad), all atomic species in PSO can be resolved in 3D at doses $\geq 10^8$ $e^-/Å^2$ with HSA $\leq$ 0.75α. For light atoms in closely spaced layers, resolvability is more dose-sensitive and HSA needs to minimize.

It is important to note that the phase value of reconstructed atomic peaks varies between different atomic species, as previously shown in literature such as $LaB_6$[50] and $MoS_2$[41]. In PSO, Pr columns exhibit the strongest phase contrast, followed by Sc, while O columns produce the weakest signal. This is consistent with trends observed in previous work[35]. This relationship remains consistent throughout the entire range of HSAs examined (Fig. 5a), reflecting it depending on atomic number and scattering strength. Therefore, peak detection algorithms can be adapted to account for species-specific contrast. As shown in Fig. 5b, the 3D localization of Pr, Sc, and O atoms can be achieved by applying element-specific detection thresholds the reconstruction at the dose of $10^8$ $e^-/Å^2$, as detailed in Supplementary Note 6. The parameters can be adjusted based on expected signal strength for each atomic species to achieve reliable full 3D atomic position reconstruction under imaging conditions of $\geq 10^8$ $e^-/Å^2$ with HSA $\leq$ 0.75α. MHP provides suitable balance between atomic-layer resolved 3D imaging and high EELS signal, enabling dose-efficient correlative analysis for practical applications that need both structure imaging and spectroscopy.

## 3. Discussion

The ability to simultaneously achieve sub-angstrom lateral resolution and atomic-layer-resolved depth sectioning is central to the development of next-generation 3D electron imaging techniques. Our results demonstrate that MHP offers a flexible, dose-dependent framework for correlative structural and chemical mapping, particularly in thick, strongly scattering materials like PSO. CSA, HSA, total electron dose and TLV together govern the fidelity of lateral and depth atomic resolution at the same collection angle.

In the low dose regime ($< 10^3$ $e^-/Å^2$), which is relevant for beam-sensitive materials, such as zeolites, solid electrolyte, organic-inorganic perovskites, MHP reconstructions remain feasible and lateral resolution is around 1 Å up to ~0.5α, while allowing ~50% of the dose to be directed to EELS. However, beyond this HSA, lateral resolution degrades rapidly (>2 Å), and noise dominates lighter atomic columns. In the intermediate dose regime ($10^4$–$10^6$ $e^-/Å^2$) applicable to systems such as lithium battery cathodes and in the high dose regime ($10^7$–$10^8$ $e^-/Å^2$) typical for functional oxides or semiconductors, lateral resolution remains better than 0.7 Å up to ~0.75-0.95α, enabling over 70% of total dose to be used for EELS. This capability supports high-resolution, high-contrast 2D structural imaging with ptychography together with correlative chemical analysis, such as elemental identification and electronic structure using core-loss EELS or energy-loss near-edge structure (ELNES) [2-4,6]. However, atomic-layer-by-layer 3D reconstruction remains challenging at these dose levels. To resolve features such as atomic layers, dopants, and defects in depth, their separation must exceed the theoretical depth resolution, typically by at least a factor two, as demonstrated in the case of PSO.

In the ultrahigh dose regime ($\geq 10^8$ $e^-/Å^2$), permitted in highly stable, radiation-tolerant samples such as oxides and metals, full atomic-layer-resolved reconstructions can be potentially obtained with the ultimate limit set by TLV. Under cryogenic conditions, both lateral and depth resolutions reach their theoretical limits of ~0.5 Å laterally and ~2.9 Å in depth across a wide HSA range up to 0.95α. In this regime, all atomic planes, including closely spaced light-element layers, can be accurately recovered without compromising compatibility with EELS. Apart from the aforementioned use in core-loss or ELNES, this regime also could provide optimal electron counts for high-quality extended energy-loss fine structure (EXELFS) analysis, which enables high signal and noise ratio measurements of bonding environments and coordination shells[51,52]. The ultrahigh dose levels in this regime provide significant advantages to EXELFS features because they require high signal-to-noise ratios to detect their subtle features.

Therefore, MHP can be adjusted to match diverse experimental goals by tuning dose, convergence angle, HAS and operating temperature. Its ability to provide atomic-layer-resolved imaging makes it particularly well-suited for investigating a wide range of structurally and chemically complex material systems, such as structure or functional materials. These include local chemical and structural disorders associated with point defects and dopants, dislocation cores and strain fields, interfacial atomic configurations in heterostructures, superlattices and grain boundaries with compositional modulation.

As the depth resolution of multislice ptychography is sensitive to lattice vibrations, the use of highly stable, low-drift *in-situ* cryo-holders[53-57] operating at liquid nitrogen, liquid helium, or variable temperatures provide essential advantages, including phonon suppression and radiation damage reduction. In addition, given the demonstrated sensitivity of ptychographic phase contrast to spin[58,59] and charge[60,61], multislice ptychography, including the MHP approach, potentially emerges as powerful new method to probe the lattice, spin, and charge degrees of freedom in three dimensions at the atomic level. Combined with simultaneous spectroscopic capabilities, this approach offers a powerful characterization platform for studying quantum phenomena[62] in condensed matter systems at low temperatures, including ferroic ordering, charge-order transitions and superconductivity.

## 4. Conclusion

In this work, we introduce a new imaging modality and reconstruction algorithm, termed MHP and validate its ability to overcome the limitations of SHP by addressing multiple scattering in thick specimens, achieving enhanced lateral and depth resolution from hollow diffraction patterns. Our theoretical analyses of probe CSA, HSA, electron dose and lattice vibrations show that MHP provides an adjustable balance between lateral resolution, depth sectioning, and EELS compatibility across various dose ranges. With up to 70% of the total dose available for EELS analysis, it can achieve sub-angstrom lateral resolution at intermediate doses ($10^4$–$10^8$ $e^-/Å^2$) and perform full 3D atomic-layer reconstruction including light-element layers at ultrahigh doses ($\geq 10^8$ $e^-/Å^2$) at the resolution limit of thermal lattice vibrations. Therefore, MHP potentially provides a powerful, tunable framework which enables correlative 3D imaging and chemical mapping in complex materials, such as energy and quantum materials. Future development of detector geometry, algorithm design, artificial intelligence and stable, and thermostability of cooling holders will further enhance the robustness and efficiency of hollow ptychography to achieve more accurate and comprehensive 3D materials analysis in electron microscopy.

**Acknowledgements**

P.W. acknowledges funding from BBSRC International Institutional Partnership Fund and the University of Warwick Research Development Fund (RDF) 2021-22 Science Development Award. Y.L. is supported by studentships within the Engineering and Physical Sciences Research Council supported Centre for Doctoral Training in Modelling of Heterogeneous Systems, Grant No. EP/S022848/1.

**Author contributions**

P.W. conceived the overall project and supervised Y.L., who carried out all the ptychographic reconstructions and relevant simulations. P.W. and Y.L. discussed the results and wrote the paper.

**Data availability**

All data that support this study are available from the corresponding author on request.

**Competing interests**

The authors declare no competing interests.

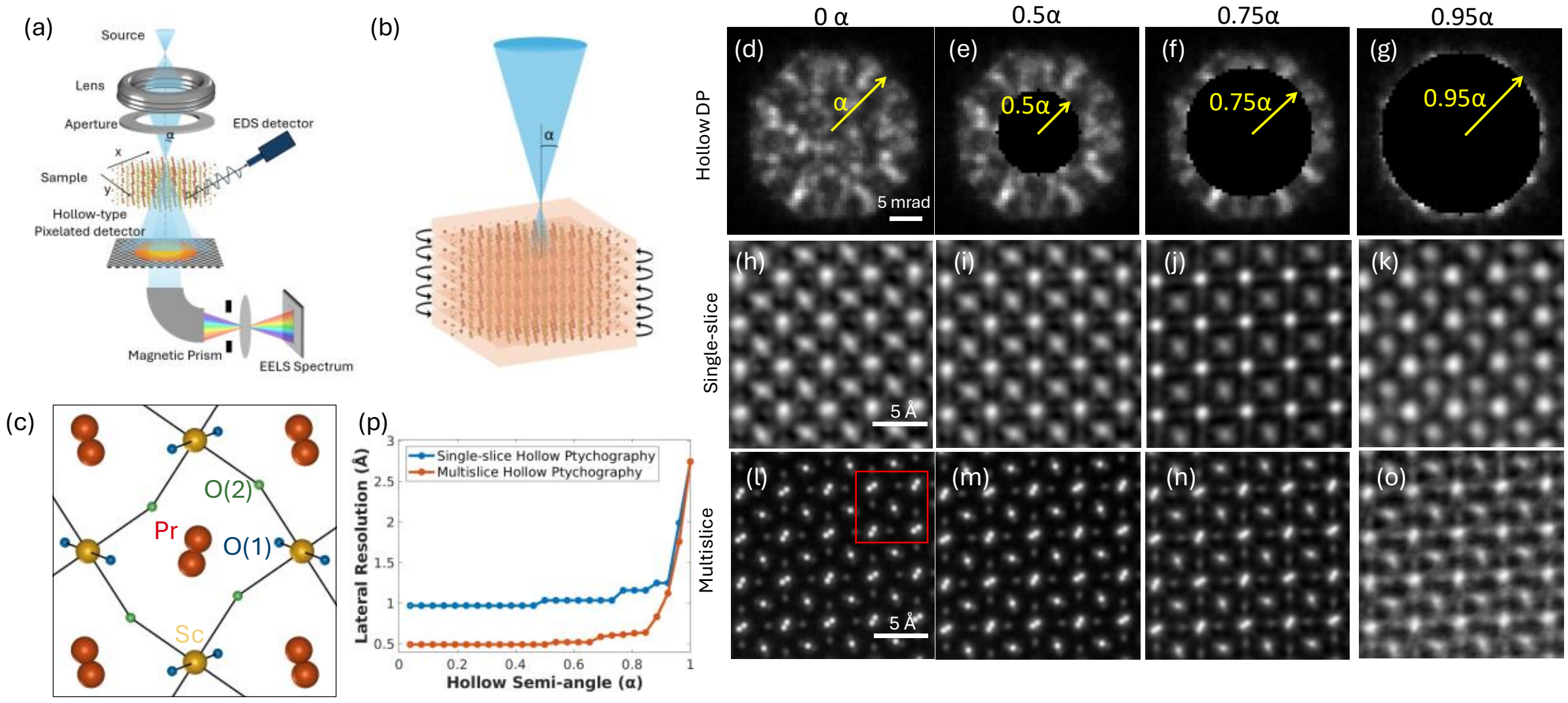

**Figure 1**: Evaluation of lateral resolution under varying hollow semi-angle (HSA) in single-slice (SHP) and multi-slice (MHP) hollow ptychography. (a) Schematic of the experimental configuration for MHP. (b) Sectioning a thick sample into a set of thin slices using a multislice approach. (c) Projected atomic model of $PrScO_3$ along the [001] direction. (d–g) Experimental hollow diffraction patterns with HSAs of 0, 0.5, 0.75, and 0.95 times the probe convergence semi-angle (CSA), α. Reconstructed phase images of $PrScO_3$ using SHP (h–k) and MHP (l-o) with corresponding diffraction patterns in (d-g). (p) Quantitative comparison of lateral resolution as a function of HSAs for SHP (—) and MHP (—).

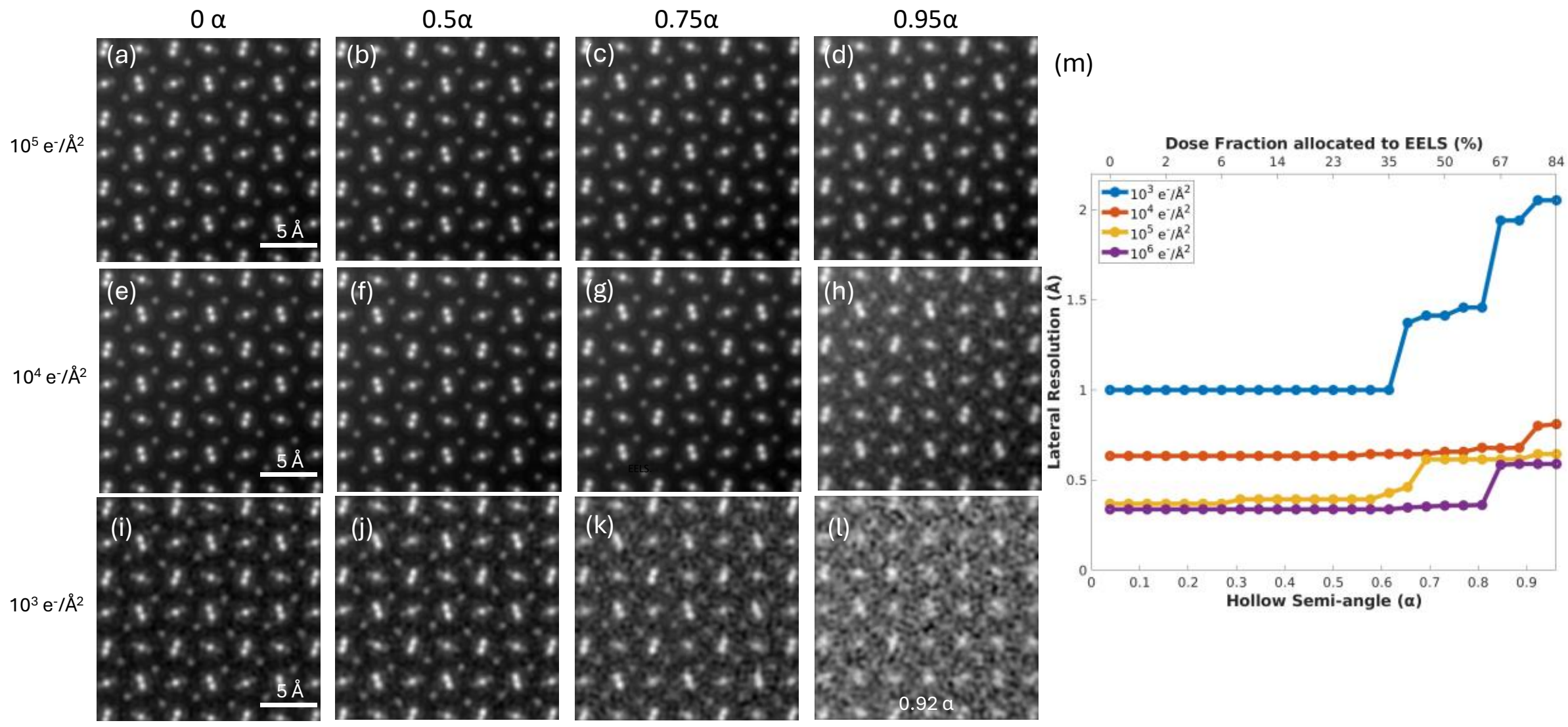

**Figure 2**: Effect of dose and HSA on simulated MHP reconstructions. Phase reconstructions at electron doses of $10^5$ (a–d), $10^4$ (e–h), and $10^3$ (i–l) e⁻/Å², with increasing HSAs from 0 to 0.95 α. (m) Lateral resolution of the reconstructed phases as a function of HSA (bottom x-axis) and the corresponding percentage of the total dose allocated to EELS acquisition (op x-axis), for doses of $10^3$ (- • -), $10^4$ (- • -), $10^5$ (- • -) and $10^6$ (- • -) e⁻/Å².

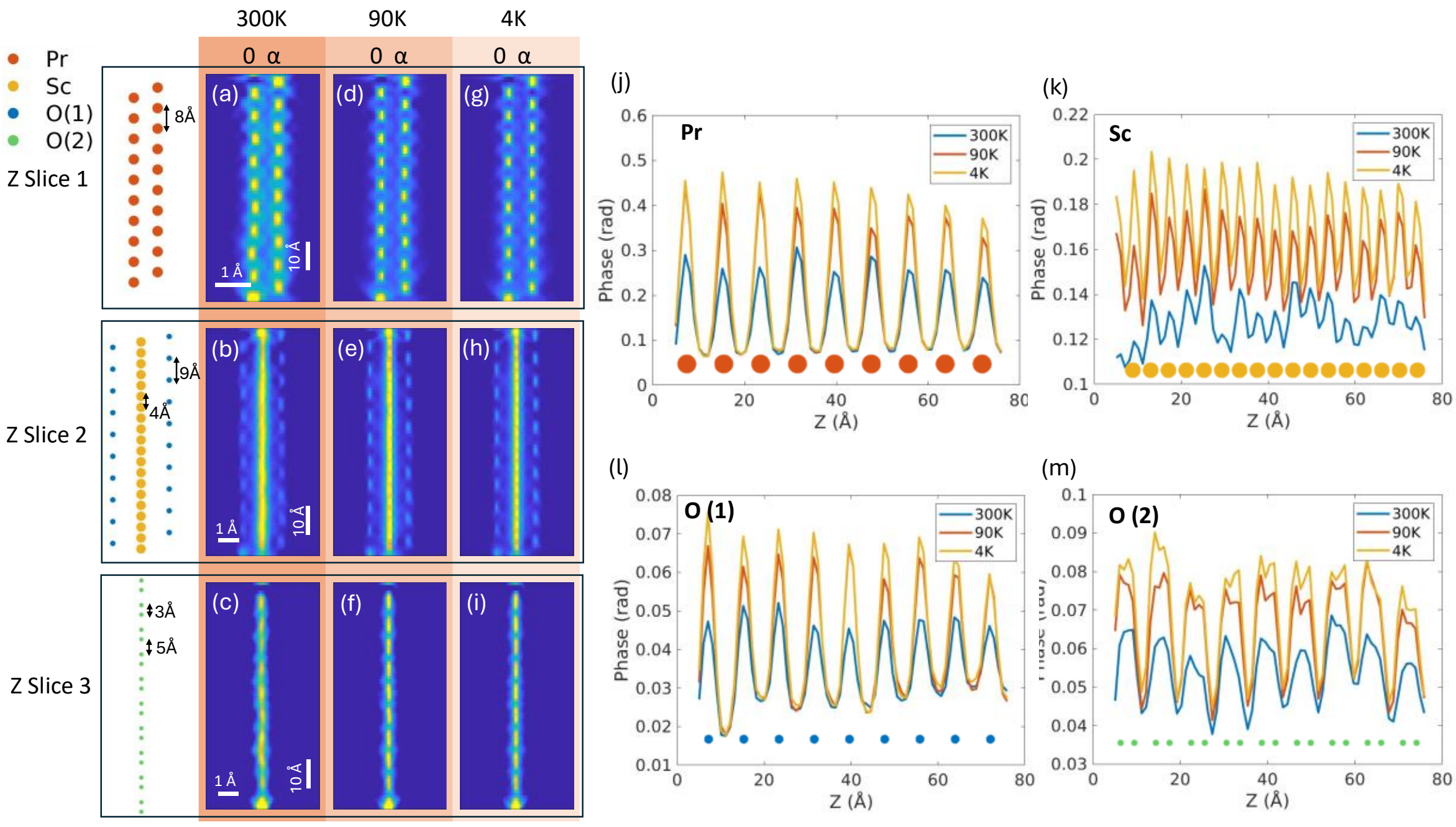

**Figure 3:** Thermal-lattice-vibration-dependent 3D MHP optical sectioning of a $PrScO_3$ crystal using full simulated diffraction pattens (HSA = 0α) at a dose of $10^8$ $e^-/Å^2$. (a-c), (d-f) and (g-i) Z-slices extracted from three representative positions (labeled 1-3 in a unit cell, as shown in Supplementary Fig. 1b) in the 3D reconstructions of $PrScO_3$ at 300 K, 90 K, and 4K, respectively. (j-m) Depth profiles extracted along atomic columns of Pr, Sc, O(1), and O(2), respectively under the different temperature conditions as shown in (a-i).

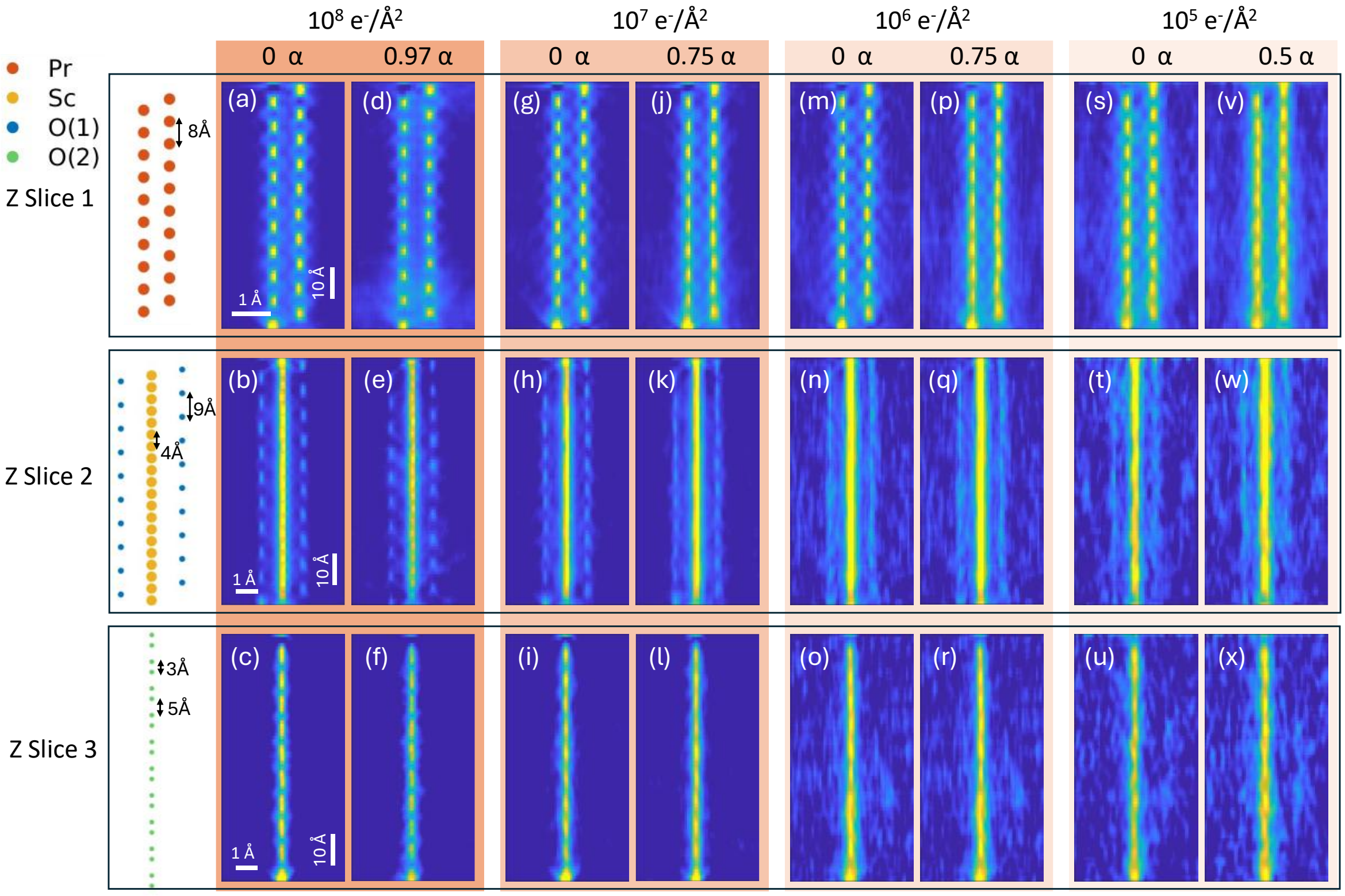

**Figure 4**: Simulated 3D MHP optical sectioning of a $PrScO_3$ crystal under varying electron dose and HSA conditions. (a-f) Z-slices extracted from three representative positions (labeled 1-3 in a unit cell, as shown in Supplementary Fig. 1b) in the 3D reconstruction of $PrScO_3$ at a dose of $10^8$ $e^-/Å^2$ with HSAs of 0α (a-c) and 0.97α (d-f). Reconstructions at lower doses: $10^7$ $e^-/Å^2$ with HSAs of 0α (g-i) and 0.75α (j-l); $10^6$ $e^-/Å^2$ with HSAs of 0α (m–o) and 0.75α (p-r); and $10^5$ $e^-/Å^2$ with HSAs of 0α (s-u) and 0.5α (v-x), respectively.

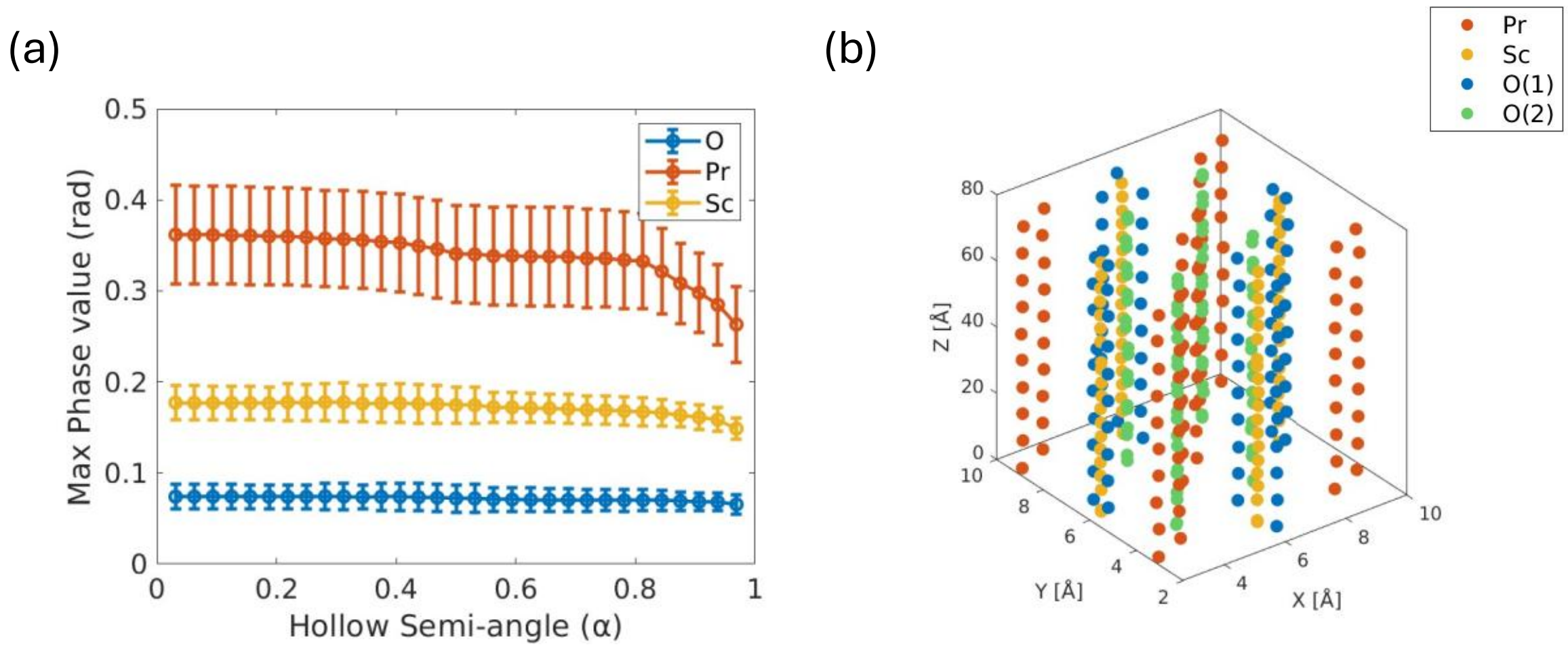

**Figure 5:** Quantitative evaluation of phase signal strength and full 3D atomic positions of $PrScO_3$. (a) Maximum phase values of Pr **(- • -)**, Sc **(- • -)**, and O **(- • -)** atoms as a function of HSA (α). Error bars indicate standard deviations across multiple measurements. (b) Reconstructed full 3D atomic positions of $PrScO_3$ visualized along x, y, and z axes.

**Supplementary Information**

# Multislice Hollow Ptychography for Simultaneous Atomic-Layer-Resolved 3D Structural Imaging and Spectroscopy

Yu Lei and Peng Wang*

## Supplementary Figures

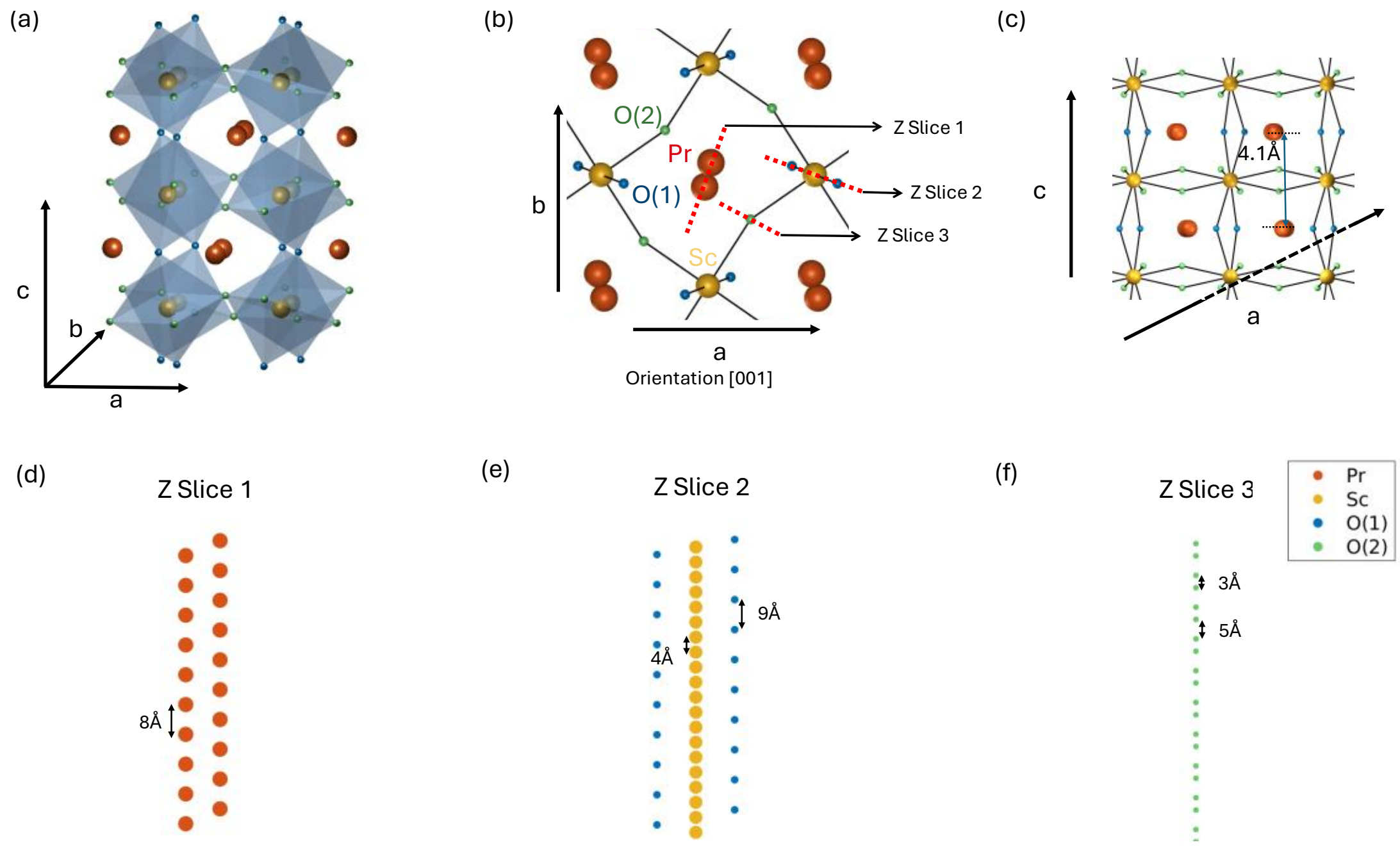

## Supplementary Figure 1

Crystal structure and atomic arrangements of $PrScO_3$ projected along different orientations. (a) Orthorhombic perovskite structure of $PrScO_3$ viewed along the [010] direction. (b) Projection of a unit cell along the [001] zone axis (c) Perspective view showing relative positions and interlayer spacings of Pr atoms. (d–f) Atomic configuration in the Z slices across the three positions indicated by red dashed-lines labeled with Z Slice 1-3, showing atomic arrangement of Pr (d), Sc and O(1) (e), and O(2) (f) atoms, respectively.

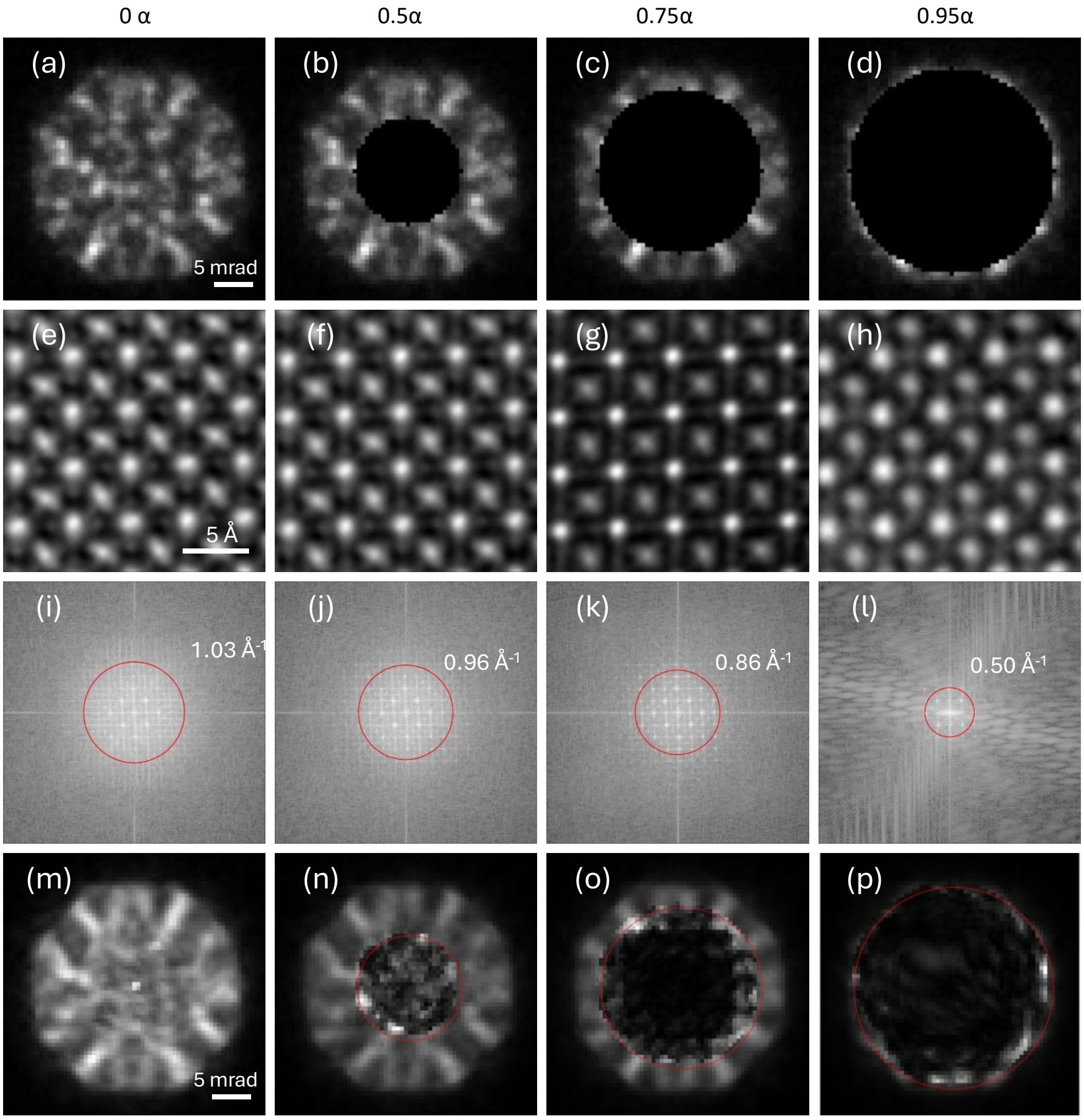

**Supplementary Figure 2**

(a-d) Hollow diffraction patterns at a dose of $1.2x10^6$ $e^-/Å^2$, with increasing hollow angles, corresponding to HSAs of 0α, 0.50α, 0.75α, and 0.95α, respectively. (e-h) Corresponding reconstructed phase images using SHP. (i-l) Corresponding power spectra of reconstructed phase images. (m-p) Diffraction patterns calculated from the resultant objects reconstructed by SHP at the corresponding hollow angles.

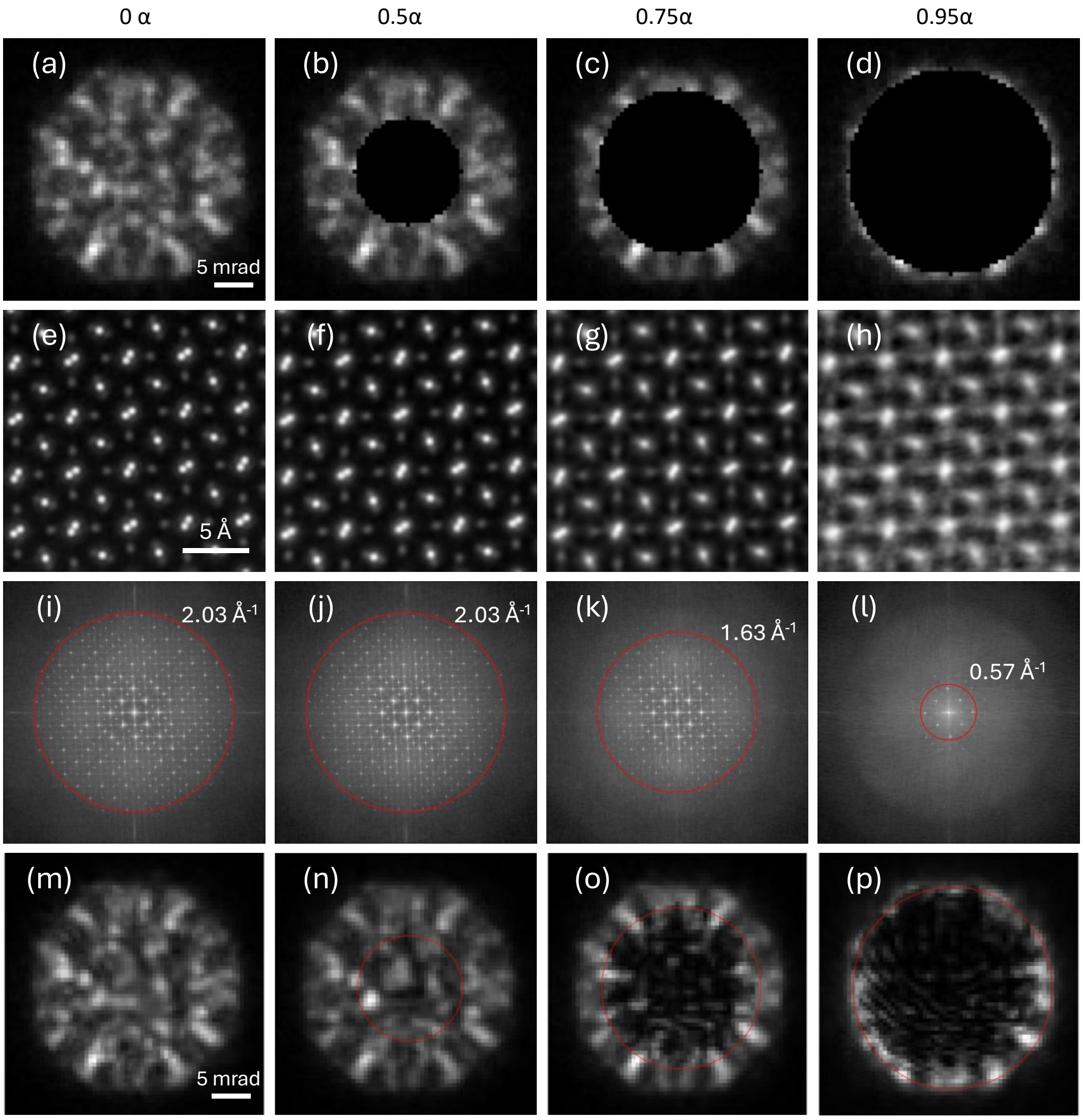

**Supplementary Figure 3**

(a-d) Hollow diffraction patterns at a dose of $1.2x10^6$ $e^-/Å^2$, with increasing hollow angles, corresponding to HSAs of 0α, 0.50α, 0.75α, and 0.95α, respectively. (e-h) Corresponding reconstructed phase images using MHP. (i-l) Corresponding power spectra of reconstructed phase images. (m-p) Diffraction patterns calculated from the resultant objects reconstructed by MHP at the corresponding hollow angles.

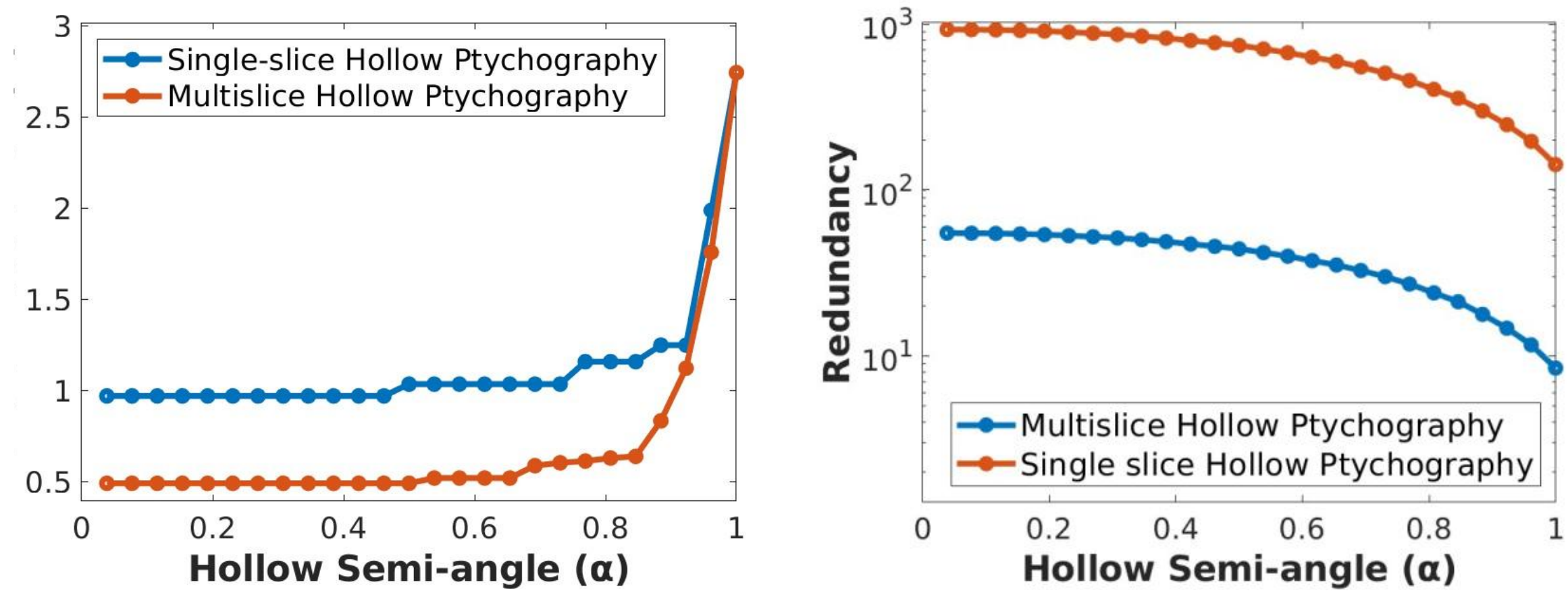

**Supplementary Figure 4**

(a-b) Quantitative comparison of lateral resolution and redundancy, respectively as a function of HSAs for SHP (—) and MHP (—).

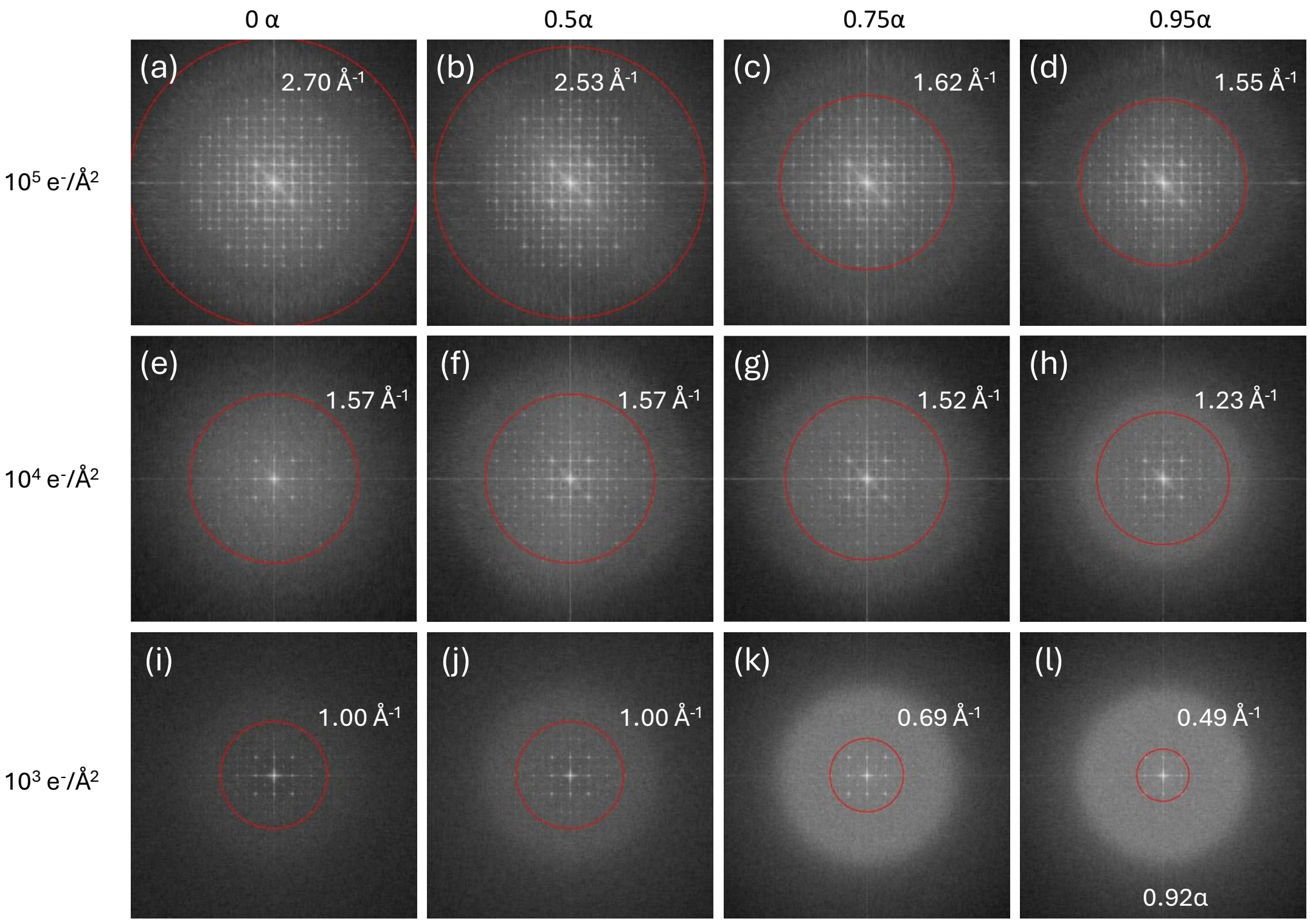

**Supplementary Figure 5**

(a–d) Power spectra of phase reconstructions at electron doses of $10^5$ (a–d), $10^4$ (e–h), and $10^3$ (i–l) $e^-/Å^2$, with increasing HSAs from 0 to 0.95α, shown in Fig. 2.

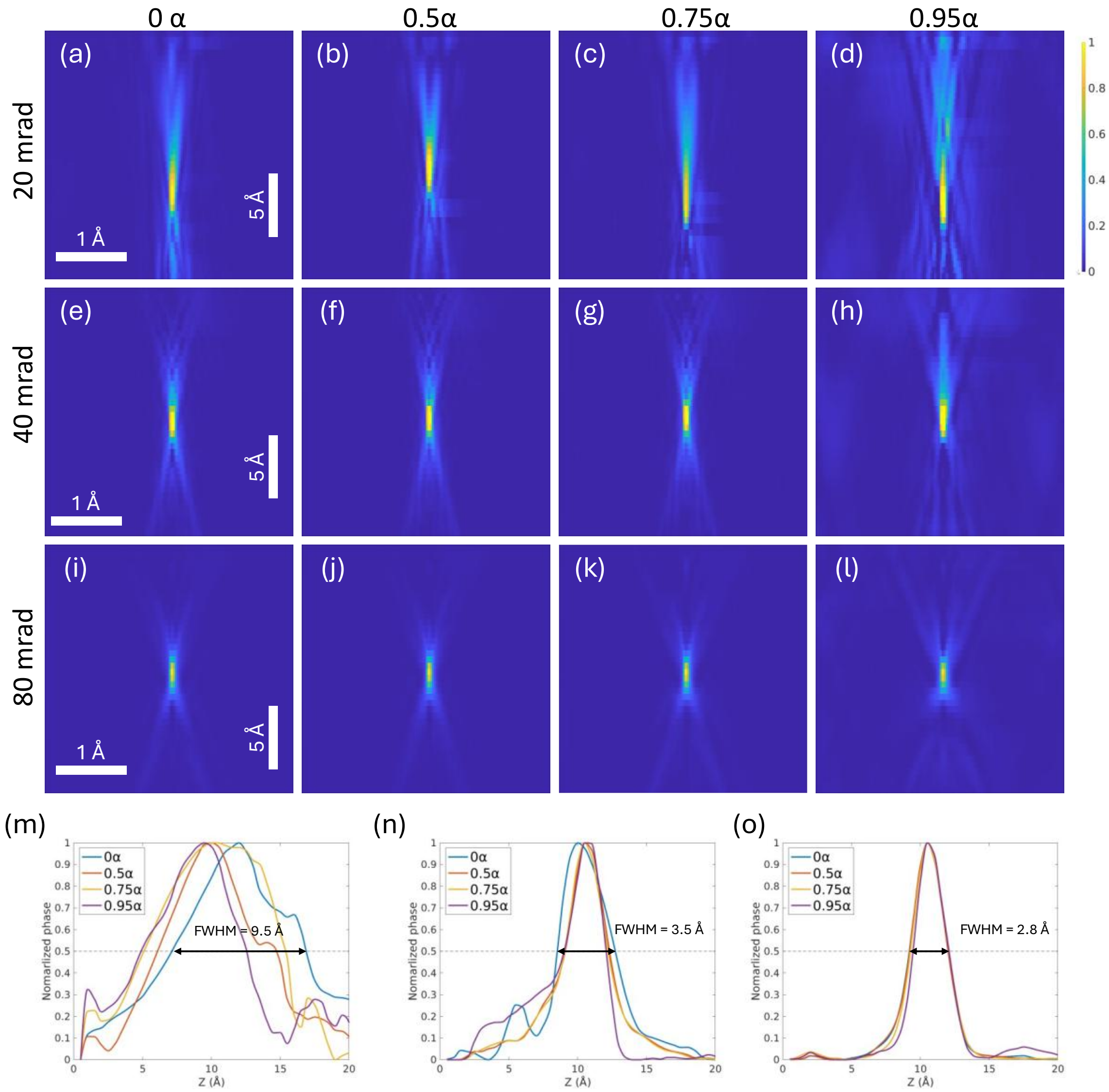

**Supplementary Figure 6**

3D MHP optical sectioning of a single Pr atom under varying probe CSAs and HSAs. MHP x-z cross sections of 3D reconstructions of the Pr atom with the probe CSAs of 20 (a-d), 40 (e-h), and 80 (i-l) mrad with increasing HSAs of 0α, 0.5α, 0.75α and 0.95α. Corresponding depth profiles across the Pr atom for the CSAs of 20 (m), 40 (n) and 80 (o) mrad.

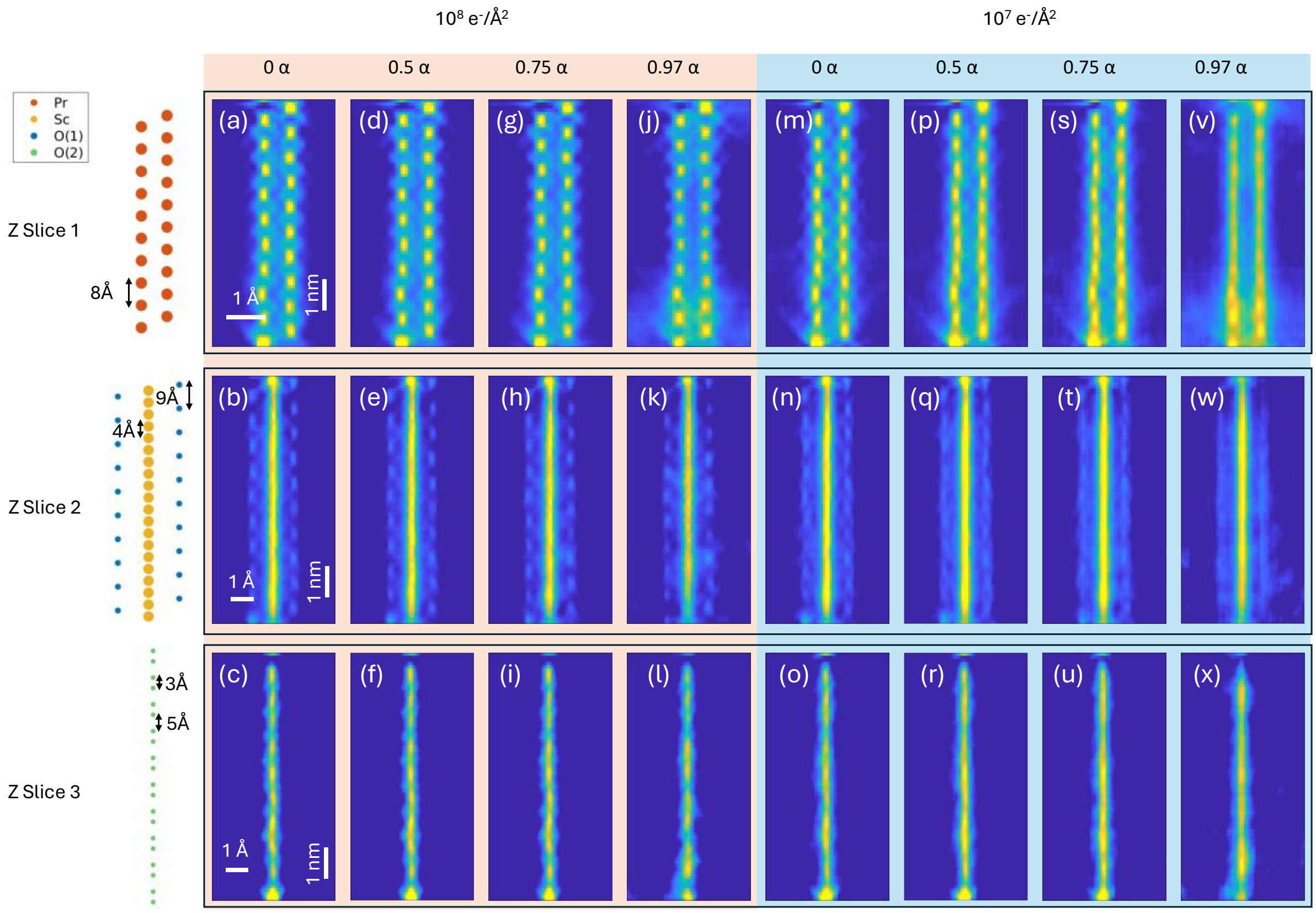

**Supplementary Figure 7**

3D MHP optical sectioning of a $PrScO_3$ crystal at 300 K under varying electron dose and HSA conditions. (a-l) Z-slices extracted from three representative positions (labeled 1-3 in a unit cell, as shown in Supplementary Fig. 1b) in the 3D reconstruction of $PrScO_3$ at a dose of $10^8$ e⁻/Å² with HSAs of 0α (a-c), 0.5α (d-f), 0.75α (g-i) and 0.97α (j-l). Reconstructions at the lower dose of $10^7$ e⁻/Å² with HSAs of 0α (m-o), 0.5α (p-r), 0.75α (s-u) and 0.97α (v-x).

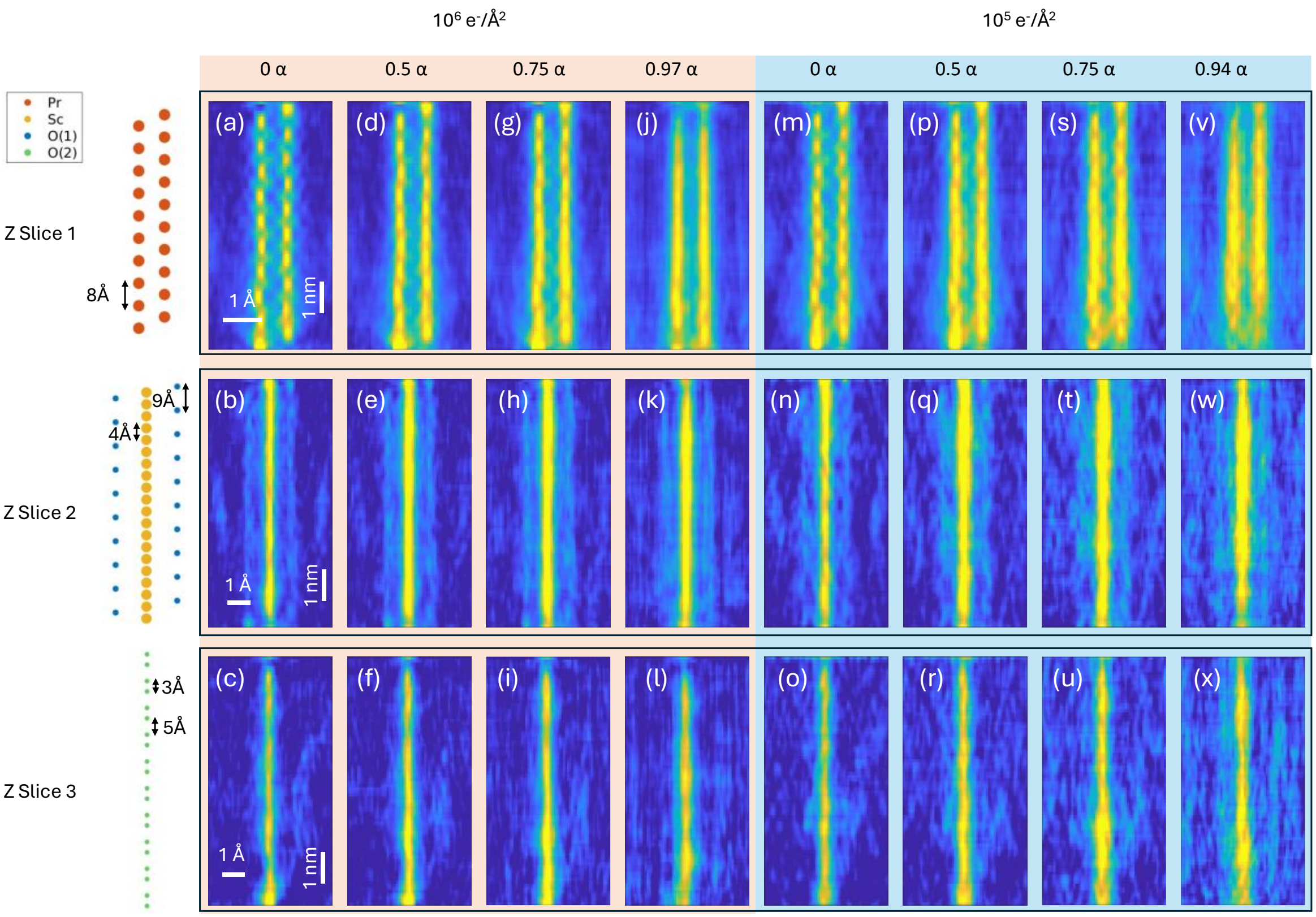

**Supplementary Figure 8**

3D MHP optical sectioning of a $PrScO_3$ crystal at 300 K under varying electron dose and HSA conditions. (a-l) Z-slices extracted from three representative positions (labeled 1-3 in a unit cell, as shown in Supplementary Fig. 1b) in the 3D reconstruction of $PrScO_3$ at a dose of $10^6$ e⁻/Å² with HSAs of 0α (a-c), 0.5α (d-f), 0.75α (g-i) and 0.97α (j-l). Reconstructions at the lower dose of $10^5$ e⁻/Å² with HSAs of 0α (m-o), 0.5α (p-r), 0.75α (s-u) and 0.97α (v-x).

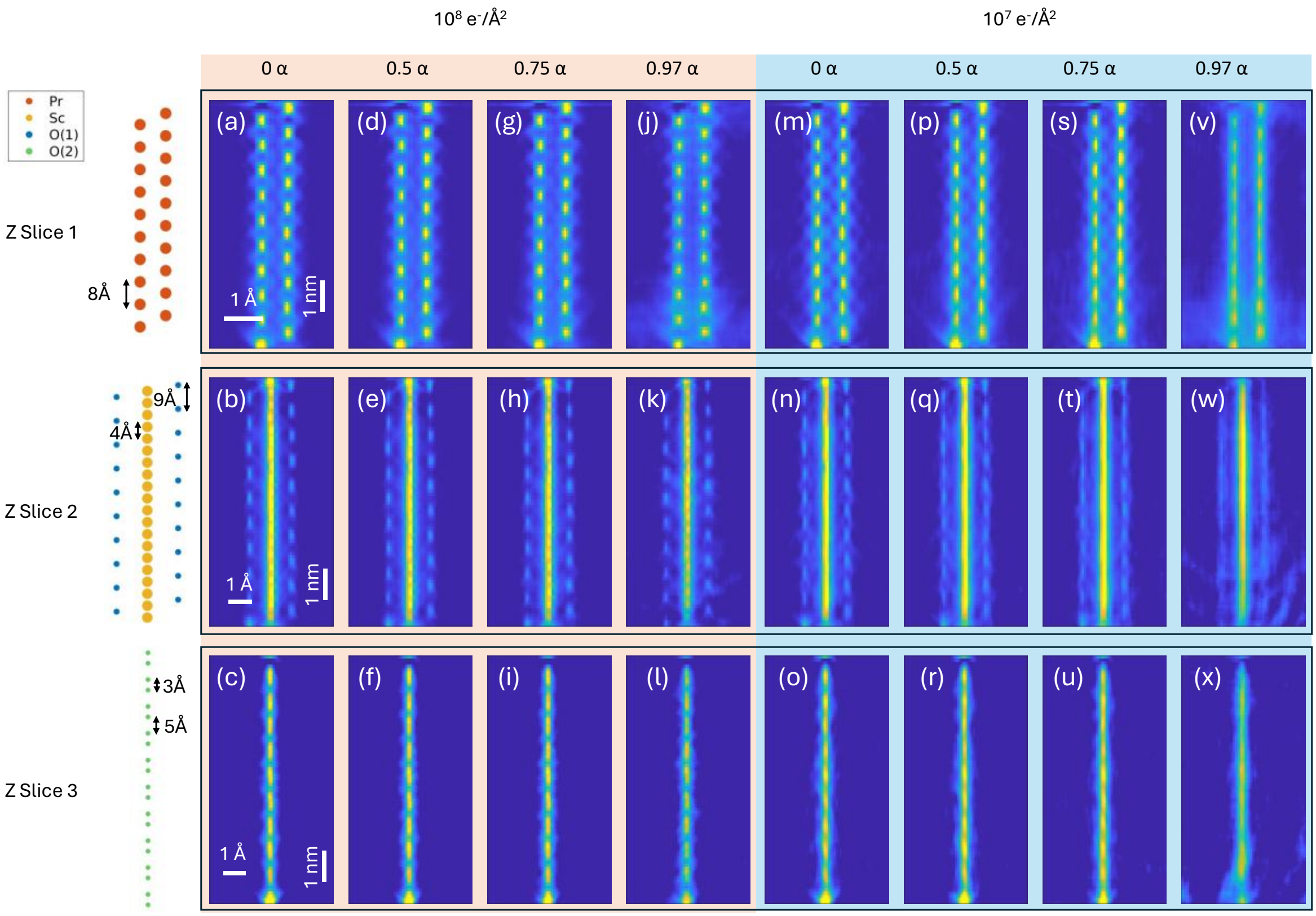

**Supplementary Figure 9**

3D MHP optical sectioning of a $PrScO_3$ crystal at 90 K under varying electron dose and HSA conditions. (a-l) Z-slices extracted from three representative positions (labeled 1-3 in a unit cell, as shown in Supplementary Fig. 1b) in the 3D reconstruction of $PrScO_3$ at a dose of $10^8$ e⁻/Å² with HSAs of 0α (a-c), 0.5α (d-f), 0.75α (g-i) and 0.97α (j-l). Reconstructions at the lower dose of $10^7$ e⁻/Å² with HSAs of 0α (m-o), 0.5α (p-r), 0.75α (s-u) and 0.97α (v-x).

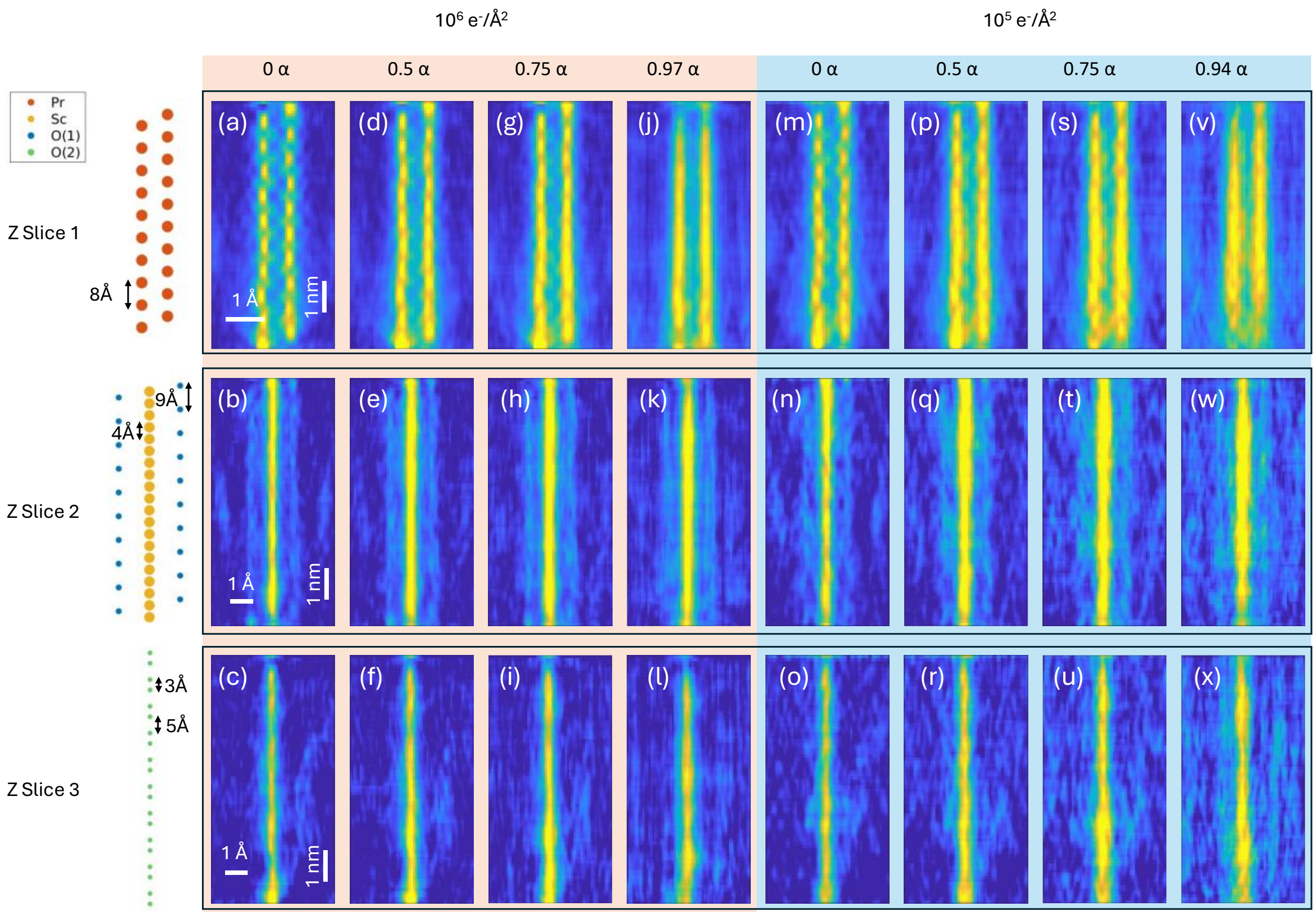

**Supplementary Figure 10**

3D MHP optical sectioning of a $PrScO_3$ crystal at 90 K under varying electron dose and HSA conditions. (a-l) Z-slices extracted from three representative positions (labeled 1-3 in a unit cell, as shown in Supplementary Fig. 1b) in the 3D reconstruction of $PrScO_3$ at a dose of $10^6$ $e^-/Å^2$ with HSAs of 0α (a-c), 0.5α (d-f), 0.75α (g-i) and 0.97α (j-l). Reconstructions at the lower dose of $10^5$ $e^-/Å^2$ with HSAs of 0α (m-o), 0.5α (p-r), 0.75α (s-u) and 0.97α (v-x).

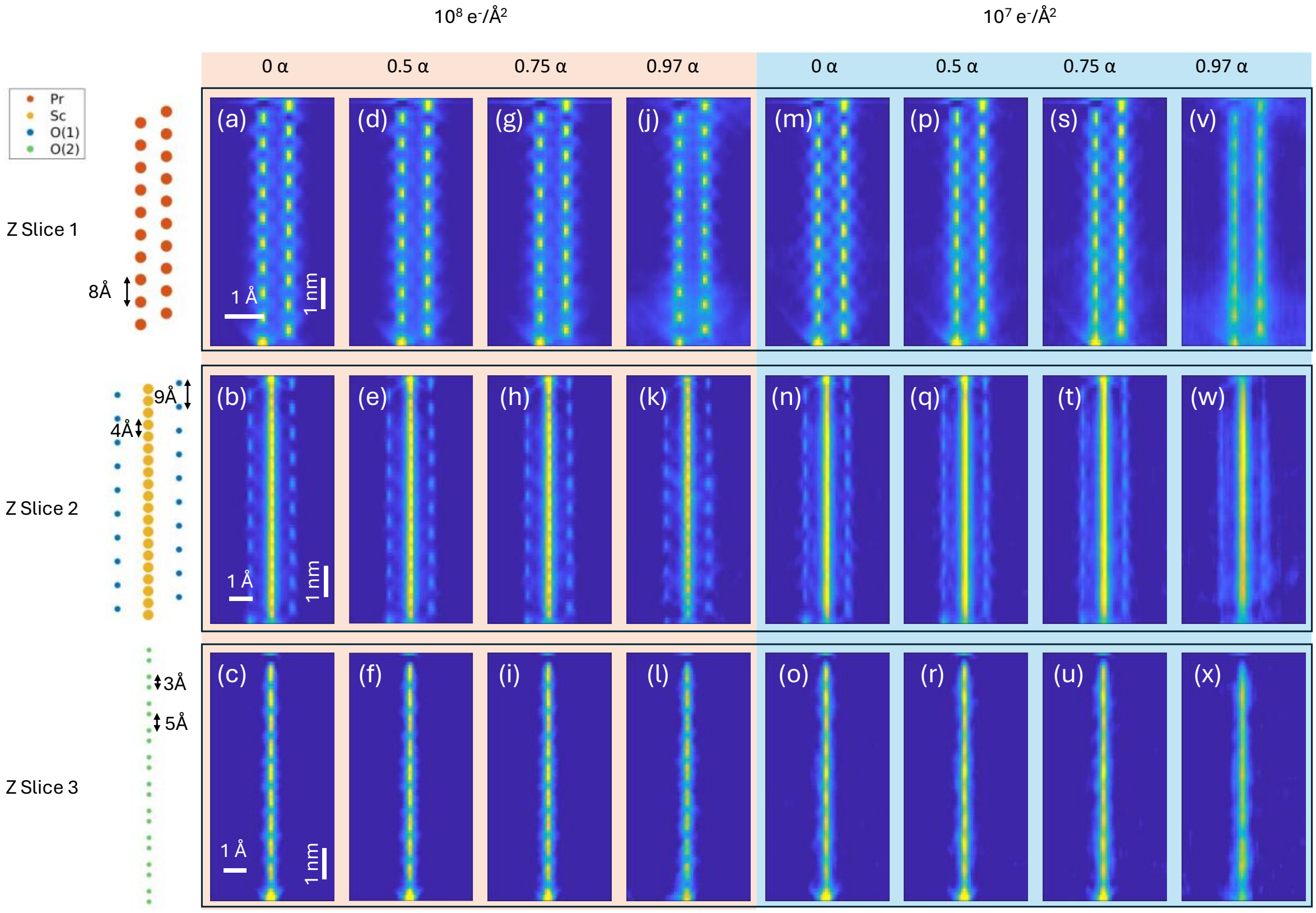

**Supplementary Figure 11**

3D MHP optical sectioning of a $PrScO_3$ crystal at 4 K under varying electron dose and HSA conditions. (a-l) Z-slices extracted from three representative positions (labeled 1-3 in a unit cell, as shown in Supplementary Fig. 1b) in the 3D reconstruction of $PrScO_3$ at a dose of $10^8$ e⁻/Å² with HSAs of 0α (a-c), 0.5α (d-f), 0.75α (g-i) and 0.97α (j-l). Reconstructions at the lower dose of $10^7$ e⁻/Å² with HSAs of 0α (m-o), 0.5α (p-r), 0.75α (s-u) and 0.97α (v-x).

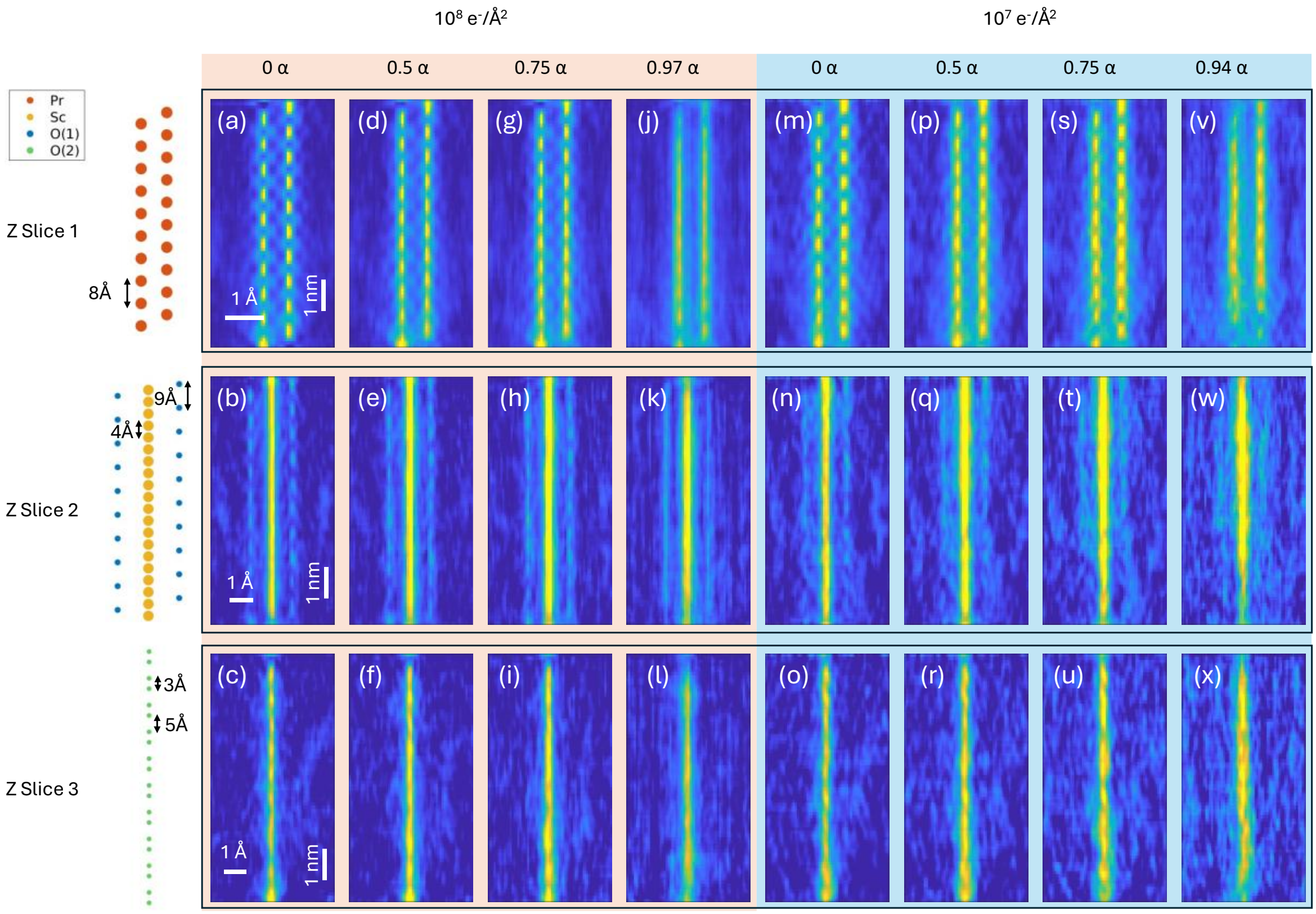

**Supplementary Figure 12**

3D MHP optical sectioning of a $PrScO_3$ crystal at 4 K under varying electron dose and HSA conditions. (a-l) Z-slices extracted from three representative positions (labeled 1-3 in a unit cell, as shown in Supplementary Fig. 1b) in the 3D reconstruction of $PrScO_3$ at a dose of $10^6$ $e^-/Å^2$ with HSAs of 0α (a-c), 0.5α (d-f), 0.75α (g-i) and 0.97α (j-l). Reconstructions at the lower dose of $10^5$ $e^-/Å^2$ with HSAs of 0α (m-o), 0.5α (p-r), 0.75α (s-u) and 0.97α (v-x).

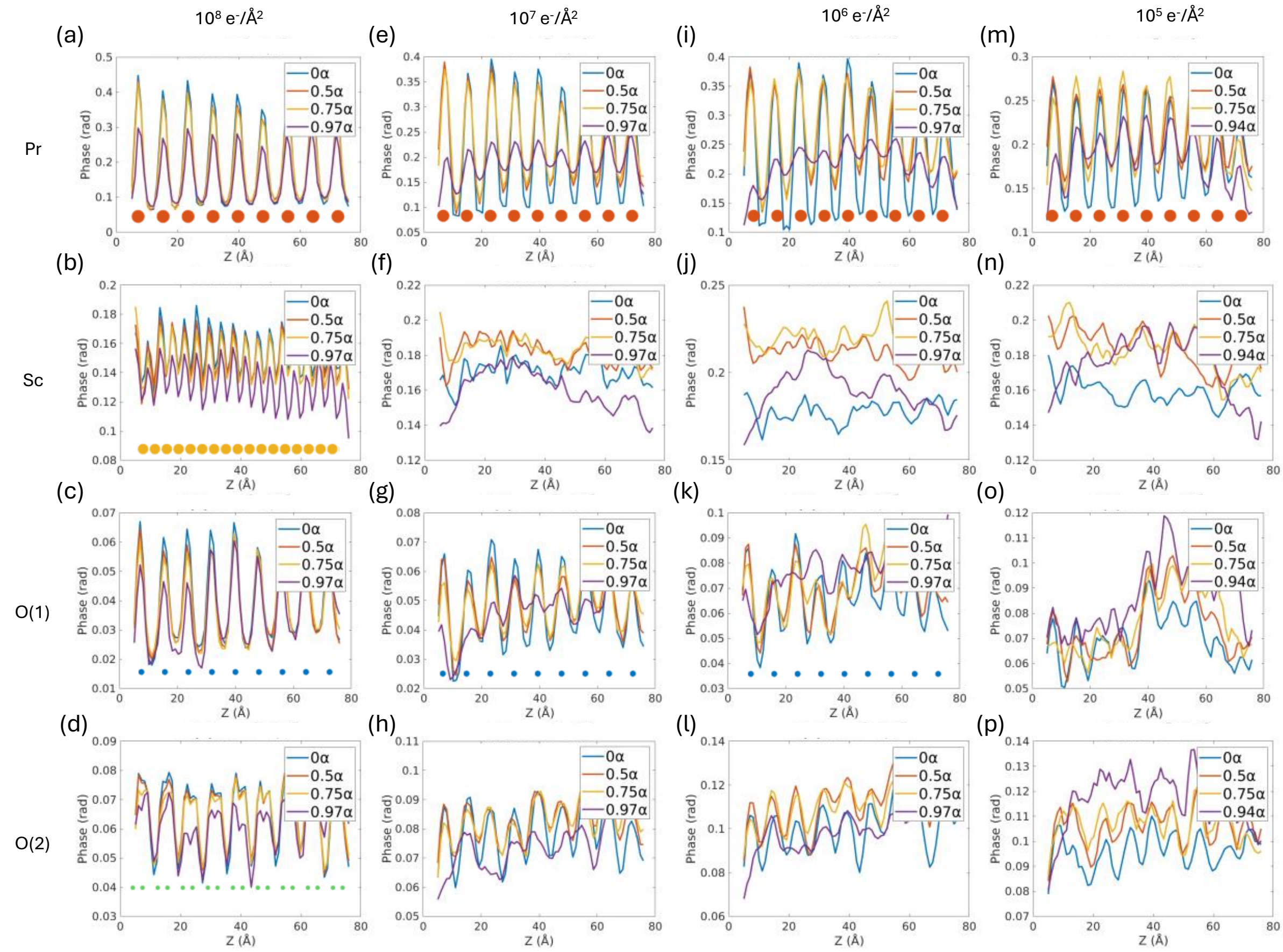

**Supplementary Figure 13**

Depth profiles along atomic columns of Pr, Sc, O(1), and O(2) in 3D MHP optical sectioning of a $PrScO_3$ crystal at 90 K with different HSAs of 0α, 0.5α, 0.75α and 0.97α at decreasing electron doses of $10^8$ (a–d), $10^7$ (e–h), $10^6$ (i–l) and $10^5$ (m–p) e⁻/Å², respectively.

**Supplementary Table 1**. **Atomic coordinates and displacement parameters[1] (in Å²)**

| Atom | Site | x | y | z | $U_{11}$ | $U_{22}$ | $U_{33}$ | $U_{12}$ | $U_{13}$ | $U_{23}$ |
|---|---|---|---|---|---|---|---|---|---|---|
| Pr(1) | 4c | 0.44930(6) | ¼ | 0.48788(6) | 0.0092(3) | 0.0097(3) | 0.0091(3) | 0 | 0.00065(9) | 0 |
| Sc(1) | 4b | 0 | 0 | ½ | 0.0077(5) | 0.0075(6) | 0.0074(6) | 0.0001(5) | 0.0003(3) | 0.0004(3) |
| O(1) | 4c | 0.395(7) | ¼ | 0.6052(8) | 0.012(2) | 0.009(2) | 0.011(2) | 0 | 0.001(1) | 0 |
| O2(2) | 8d | 0.1992(5) | 0.0555(4) | 0.1977(5) | 0.010(1) | 0.014(2) | 0.010(1) | 0.002(1) | 0.003(1) | 0.001(3) |

**Supplementary Table 2. Simulation parameters for dose-dependent multi-slice hollow ptychography**

| Dataset | Setting (1) | Setting (2) | Setting (2) |
|---|---|---|---|
| Sample | $PrScO_3$ | $PrScO_3$ | $PrScO_3$ |
| Accelerating voltage (**kV**) | 300 | 300 | 300 |
| Convergence semi-angle ($\mathbf{mrad}$) | 21.4 | 21.4 | 21.4 |
| Maximum detection semi-angle ($\mathbf{mrad}$) | 105 | 105 | 105 |
| Defocus ($\mathbf{nm}$) | -2.4 | -2.4 | -2.4 |
| Diffraction pattern dimension ($\mathbf{pixels}$) | 256×256 | 256×256 | 256×256 |
| Raster scan dimension | 81×81 | 81×81 | 81×81 |
| Dose ($\mathbf{e^-/Å^2}$) | $10^5$ | $10^4$ | $10^3$ |
| Total thickness (**nm**) | 21 | 21 | 21 |

**Supplementary Table 3. Simulation parameters for single atom optical sectioning using multi-slice hollow ptychography**

| **Dataset** | **Setting (1)** | **Setting (2)** | **Setting (2)** |
|---|---|---|---|
| Sample | Pr | Pr | Pr |
| Accelerating voltage (**kV**) | 300 | 300 | 300 |
| Convergence semi-angle (**mrad**) | 20 | 40 | 80 |
| semi-angle (**mrad**) | 150 | 150 | 150 |
| Defocus (**nm**) | 0 | 0 | 0 |
| Diffraction pattern dimension (**pixels**) | 256×256 | 256×256 | 256×256 |
| Raster scan dimension | 160×160 | 160×160 | 160×160 |
| Dose (**$e^-$/Å$^2$**) | $10^{12}$ | $10^{12}$ | $10^{12}$ |
| Total thickness (**nm**) | 4 | 4 | 4 |

**Supplementary Table 4. Simulation parameters for multi-slice hollow ptychography optical Sectioning of a PSO Crystal**

| Dataset | Setting (1) | Setting (2) | Setting (3) | Setting (4) |
|---|---|---|---|---|
| Sample | $PrScO_3$ | $PrScO_3$ | $PrScO_3$ | $PrScO_3$ |
| Accelerating voltage (**kV**) | 300 | 300 | 300 | 300 |
| Convergence semi-angle (**mrad**) | 80 | 80 | 80 | 80 |
| Maximum detection semi-angle (**mrad**) | 160 | 160 | 160 | 160 |
| Defocus (**nm**) | -0.7 | -0.7 | -0.7 | -0.7 |
| Diffraction pattern dimension (**pixels**) | 256×256 | 256×256 | 256×256 | 256×256 |
| Raster scan dimension | 176×176 | 176×176 | 176×176 | 176×176 |
| Dose ($e^-/Å^2$) | $10^8$ | $10^7$ | $10^6$ | $10^5$ |
| Total thickness (**nm**) | 8 | 8 | 8 | 8 |

**Supplementary Note 1: Experimental 4D-STEM dataset of $PrScO_3$**

An experimental 4D-STEM ptychographic dataset of a 20 nm thick $PrScO_3$ (PSO) sample, previously reported in the literature[2] was used in this study. The $PrScO_3$ single crystal was grown via the Czochralski melt growth technique and a cross-sectional TEM specimen with the thinnest region below 10 nm was prepared from the single crystal using a focused ion beam.

4D-STEM data were acquired using an aberration-corrected STEM (Titan Themis, Thermo Fisher Scientific) operated at 300 keV, equipped with a probe convergence semi-angle of 21.4 mrad. The dataset was recorded along the [001] zone axis using a pixelated detector (EMPAD, 128 × 128 pixels). A ~20 nm overfocused probe was used. Scanning was performed over a 64 × 64 grid with a step size of 0.41 Å, covering an area of 2.6 × 2.6 $nm^2$. The beam current was ~33 pA, with a dwell time of 1 ms and a readout time of 0.86 ms per scan point, giving a total electron dose of approximately $1.2 \times 10^6$ $e \cdot Å^{-2}$.

Further details on the sample, experimental setup and datasets can be found in the Reference[2].

**Supplementary Note 2: Reconstruction algorithm for multi-slice hollow ptychography**

The reconstruction in this work is based on an iterative Least Squares Maximum Likelihood (LSQML) method, modified from the Ptychoshelves framework[3-7]. The algorithm alternates between simulating the forward model and refining the object and probe estimates based on a masked likelihood-based error metric.

At each iteration, the algorithm computes the forward exit wave function $\psi$ by multiplying the current estimates of the object $O$ and the probe $P$, projected at a given scan position. This exit wave is propagated to the detector plane to obtain the predicted diffraction amplitude $\Psi$, which is compared to the experimentally measured diffraction pattern $I$. The discrepancy between $\Psi$ and $I$ is used to compute a residual wavefield $\chi$, The residual wavefield $\chi$ is computed as the difference between the current exit wave and a modified wavefield obtained by replacing the simulated diffraction amplitude with the measured one,

$$\chi = \left(\mathcal{F}^{-1}(|\sqrt{I}|\cdot\frac{|\Psi|}{\Psi}) - \psi\right)\cdot M, \tag{1}$$

and is further masked to exclude missing regions in the diffraction pattern, which serves as the gradient signal for updating both the object and the probe.

To prevent unreliable updates due to missing information in the diffraction patterns (such as central hollow regions), a binary mask $M$ is applied during the modulus constraint step. This ensures that the residual $\chi$ is computed only in the measured regions, and that no updates are driven by missing information.

The update directions for the object and probe are computed in the real space as:

$$\Delta O = \chi \cdot P^*, \tag{2}$$

$$\Delta P = \chi \cdot O^*, \tag{3}$$

where $*$ denotes the complex conjugation. These directions are then refined using overlap-consistency constraints and a least-squares criterion to determine optimal update step sizes $\beta_O$ and $\beta_P$. The refined step sizes can be obtained either by solving a local linear system or by a simplified projection method. The final updates are applied as:

$$O \leftarrow O + \beta_O \cdot \Delta O, \tag{4}$$

$$P \leftarrow P + \beta_P \cdot \Delta P, \tag{5}$$

In the case of multilayer objects, the residual wavefield is back-propagated layer by layer, and the object updates are applied separately for each layer. The probe is updated only from the first layer to maintain stability.

Optionally, the algorithm includes mechanisms for refining experimental parameters such as probe position, detector rotation, and Fourier space misalignment. Furthermore, a momentum acceleration scheme may be employed, where update directions from previous iterations are accumulated to improve convergence.

By iteratively applying this masked error-driven update process, the LSQML algorithm effectively recovers both the object and the probe function, even in the presence of missing hollow information in the diffraction patterns.

**Supplementary Note 3: Simulation parameters for dose-dependent multi-slice hollow ptychography**

**Atomic model**

The crystal model as shown in Supplementary Fig. 1. is based on the experimentally refined orthorhombic perovskite structure of $PrScO_3$[1], with lattice parameters and atomic positions listed in Supplementary Tab. 1. This structure was used to construct the simulation supercell with a slab thickness of 20 nm, corresponding to approximately 26 unit cells along the beam direction.

**4D-STEM ptychographic datasets simulations and reconstructions**

Simulated 4D-STEM datasets were calculated using abTEM[8]. A slab model of orthorhombic $PrScO_3$ was constructed along the [001] zone axis based on the atomic coordinates listed in Supplementary Tab. 1. Simulation parameters are shown in Supplementary Tab. 2. Simulations were performed at an accelerating voltage of 300 keV using a 21.4 mrad probe-forming convergence semi-angle (α). A 4D-STEM scan grid of 81 × 81 was applied, and each scan point produced a diffraction pattern on a 256 × 256 pixel array. Poisson noise was added to the simulated diffraction patterns to match experimental dose conditions, and a hollow mask was applied with variable hollow angles (e.g., 0, 0.5α, 0.75α, 0.95α) after applying Poisson noise.

All reconstructions were performed using the PtychoShevles[3] package modified to support multislice hollow ptychography and run for 400 iterations.

**Supplementary Note 4: Simulation parameters for 3D multi-slice hollow ptychography optical sectioning of a single atom**

Simulated 4D-STEM datasets were calculated using abTEM. A single Pr atom was used to build the atomic model. Simulation parameters are shown in Supplementary Tab. 3. Simulations were performed at an accelerating voltage of 300 keV using 20, 40 and 80 probe-forming convergence semi-angles (α). A 4D-STEM scan grid of 160×160 positions was applied, and each scan point produced a diffraction pattern on a 512 × 512 pixel array. Poisson noise was added to the simulated diffraction patterns at a dose of $10^{12}$ $e^-/Å^2$ and a mask was applied with variable hollow angles (e.g., 0, 0.5α, 0.75α, 0.95α) after applying Poisson noise.

All reconstructions were performed using the modified PtychoShevles package for 80 iterations.

**Supplementary Note 5: 3D MHP optical sectioning of a single atom**

The capability to resolve the 3D spatial arrangement of atoms within materials along the beam direction has been a long-standing goal for optical sectioning techniques in STEM, such as HAADF[9,10] and more recently, ptychography. Here, a single atom in vacuum as a point scatter is used to theoretically evaluate and compare the depth resolution limits of MHP with different probe convergence semi-angles (CSAs) while maintaining constant collection angle and total electron dose (Full simulation details are provided in Supplementary Note 4). Simulations were performed with probe CSAs of 20 (Supplementary Fig. 6a-d), 40 (Supplementary Fig. 6e-h), and 80 (Supplementary Fig. 6i-l) mrad under a dose of $10^{12}$ $e^{-}/Å^{2}$ as HSA is varied from 0, 0.5α, 0.75α, and up to 0.95α. For a probe with 20mrad convergence angle, the FWHM of the reconstructed Pr atom remains at 9.5 Å across different HSAs (Supplementary Fig. 6m). However, beyond HSA = 0.95α, reconstruction becomes unstable, with a small peak appearing in the top five z-slices. In contrast, for the 40 mrad probe, the depth resolution significantly improves, with a stable FWHM of 3.5 Å across HSA values up to 0.95α (Supplementary Fig. 6n). To explore the feasibility of atom-resolved depth sectioning, simulations using a CSA of 80 mrad were conducted. The depth reconstructions (Supplementary Fig. 6i-l) and corresponding profiles (Supplementary Fig. 6o) demonstrate a FWHM of 2.8 Å with a complete diffraction pattern dataset. As the HSA increases up to 0.95α, the depth resolution remains stable at 2.9 Å. However, beyond HSA = 0.95α, the reconstruction becomes unstable. These results show that the depth resolution in MHP is primarily determined by the CSA, provided the collection angle and total dose are constant. The depth resolution remains largely independent of the hollow geometry until extreme values (HSA > 0.95α) are reached. Increasing CSA narrows the depth of focus and enhances sensitivity to z-localized phase contrast, which results in improving depth resolution.

**Supplementary Note 6: Simulation parameters for 3D multi-slice hollow ptychography optical sectioning of a $PrScO_3$ crystal**

Simulated 4D-STEM datasets were calculated using abTEM. A slab model of orthorhombic $PrScO_3$ was constructed along the [001] zone axis based on the atomic coordinates listed in Supplementary Tab. 1. Simulation parameters are shown in Supplementary Tab. 4. Simulations were performed at an accelerating voltage of 300 keV using an 80 mrad probe-forming convergence semi-angle (α). A 4D-STEM scan grid of 81 × 81 was applied, and each scan point produced a diffraction pattern on a 256 × 256 pixel array using multislice procedures[11-13]. Poisson noise was added to the simulated diffraction patterns at a dose of $10^8$ $10^7$ $10^6$ and $10^5$ $e^-/Å^2$ and a mask was applied with variable hollow angles (e.g., 0, 0.5α, 0.75α, 0.95α) after applying Poisson noise.

Thermal atomic vibrations were accounted for using the frozen phonon approach, implemented via atomic configuration averaging. Anisotropic displacement parameters (ADPs) obtained from experimental crystallographic refinement (Supplementary Table 2) were converted to isotropic root-mean-square displacements ($\sigma$) using:

$$\sigma = \sqrt{\frac{1}{3}(U_{11} + U_{22} + U_{33})} \ , \tag{5}$$

where $U_{ii}$ are the principal components of the ADP tensor in $Å^2$ [14,15]. The calculated σ values at room temperature (300 K) were 0.0966 Å for Pr, 0.0868 Å for Sc, and 0.103 Å for O(1) and O(2), respectively.

For low-temperature simulations, these values were scaled by $\sqrt{T/300}$ to account for reduced thermal motion at 90 K (liquid nitrogen) and 4 K (liquid helium)[15,16]. At 90 K, $\sigma$ values were reduced to 0.053 Å (Pr), 0.048 Å (Sc), and 0.057 Å (O); at 4 K, a zero-point motion approximation was applied, assuming 23% of the room-temperature values, yielding $\sigma$ values of 0.022 Å (Pr), 0.020 Å (Sc), and 0.024 Å (O).

Atomic-layer 3D reconstruction using MHP of a crystalline sample at a large CSA of 80 mrad was investigated. The same $PrScO_3$ crystal model as shown in Supplementary Figure 1a was employed here shown in with a defocus of 0.7 nm applied during data acquisition. This defocus corresponds to 1 Å diameter of illumination area per probe position, consistent with the spatial resolution achievable by chemical analytical techniques such as EELS, while avoiding defocus-induced depth blurring. All reconstructions were performed using the modified PtychoShevles package for 160 iterations.

**Supplementary Note 7: Peak detection algorithms for species-specific full atomic positions**

To accurately determine atomic positions from the reconstructed 3D phase volumes, we implemented a multi-step refinement procedure, as described below:

1. Initial Atomic Model Preparation

The starting atomic model was imported from an '.xyz' file containing the atomic species and their corresponding Cartesian coordinates. To align the model with the reconstructed phase volume, the atomic structure was first translated so that its geometric center coincided with the origin. The processed coordinates served as the initial reference positions.

2. Loading and Processing the Reconstructed Phase Volume

The reconstructed 3D complex object was obtained from the final iteration (e.g., 'Niter60.mat') of the ptychographic reconstruction. The wrapped phase values of each 2D slice were extracted and stacked to form a 3D phase volume.

3. Initial Atom Refinement by Local Maximum Search

Each atom's position was refined using a coarse-to-fine localization strategy:

A small 3D sub-volume was extracted around the expected atomic position. The size of this window was determined by the element-specific interaction radius (typically 0.2 Å in-plane and 1 Å along the Z-axis). Within this volume, the voxel with the highest phase intensity was located as an initial estimate of the atomic center. A slightly larger neighborhood (e.g., 0.3 Å in-plane, 1–3 Å along Z) was then used to compute the intensity-weighted centroid of the phase volume. This resulted in sub-pixel accuracy in determining the atomic coordinates. Element-specific fitting parameters were used to account for differences in scattering strength and visibility among different atom types (e.g., Pr, Sc, O).

4. Alignment by Procrustes Analysis

To correct for residual global translation or rotation between the reconstructed positions and the reference model, a rigid Procrustes alignment was performed. This alignment preserved the scale and minimized the overall displacement between the two sets of coordinates. After alignment, the refinement procedure was repeated once to further improve the consistency and accuracy of atomic position determination.

**Supplementary References**